\documentclass{aa}  
\usepackage{graphicx}
\usepackage{txfonts}
\usepackage{natbib} 
\usepackage{subfig}

\begin{document}
   \title{The evolved circumbinary disk of AC Her: a radiative transfer, interferometric, and mineralogical study  
   \thanks{Based on observations made with ESO Telescopes at the La Silla Paranal Observatory under program ID 075.D-0605.}} 
   \author{M. Hillen\inst{1}
          \and
          B.L. de Vries\inst{2,3}
          \and
          J. Menu\inst{1}
          \and
          H. Van Winckel\inst{1} 
          \and
          M. Min\inst{4}
          \and
          G. D. Mulders\inst{5}
          }

   \institute{Instituut voor Sterrenkunde (IvS), KU Leuven,
              Celestijnenlaan 200D, B-3001 Leuven, Belgium\\
              \email{Michel.Hillen@ster.kuleuven.be}
            \and
              AlbaNova University Centre, Stockholm University, Department of Astronomy, SE-106 91, Stockholm, Sweden    
            \and
              Stockholm University Astrobiology Centre, SE-106 91 Stockholm, Sweden
            \and
              Sterrenkundig Instituut Anton Pannekoek, University of Amsterdam, Science Park 904, 1098 XH Amsterdam, The Netherlands
            \and
              Lunar and Planetary Laboratory, The University of Arizona, 1629 E. University Blvd., Tucson AZ 85721, USA
             }

   \date{Received 19 November, 2014; accepted 11 March, 2015}
   \authorrunning{Hillen et al.}
   \titlerunning{Radiative transfer models of AC\,Herculis}

 
  \abstract
   {Many post-asymptotic giant branch (post-AGB) stars in binary systems have an infrared (IR) excess arising from 
   a dusty circumbinary disk. The disk formation, current structure, and further evolution are, however, poorly understood.}
   {We aim to constrain the structure of the circumstellar material around the post-AGB binary and RV Tauri pulsator AC\,Her. 
   We want to constrain the spatial distribution of the amorphous and of the crystalline dust. 
   }
   {We present very high-quality mid-IR interferometric data that were obtained with the MIDI/VLTI instrument. We analyze 
   the MIDI visibilities and differential phases in combination with the full spectral energy distribution (SED), 
   using the MCMax radiative transfer code,
   to find a good structure model of AC\,Her's circumbinary disk. We include a grain size distribution and midplane settling of dust 
   self-consistently in our models. The spatial distribution of crystalline forsterite in the disk is investigated 
   with the mid-IR features, the 69~$\mu$m band and the 11.3~$\mu$m signatures in the interferometric data.}
   {All the data are well fitted by our best model. The inclination and position angle of the disk are precisely determined 
   at $i=50\pm8^\circ$ and $PA=305\pm10^\circ$. We firmly establish that the inner disk radius is about an order of magnitude larger
   than the dust sublimation radius. The best-fit dust grain size distribution shows that significant grain growth has
   occurred, with a significant amount of mm-sized grains now being settled to the midplane of the disk. 
   A large total dust mass $\geq10^{-3}$~M$_\odot$ is needed to fit the mm fluxes. By assuming $\alpha_{\rm{turb}}=0.01$, 
   a good fit is obtained with a small grain size power law index of 3.25, combined with a small gas/dust ratio $\leq$10. 
   The resulting gas mass is compatible with recent estimates employing direct gas diagnostics. 
   The spatial distribution of the forsterite is different from the amorphous dust, as more warm forsterite is needed 
   in the surface layers of the inner disk.}
   {The disk in the AC\,Her system is in a very evolved state, as shown by its small gas/dust ratio and large inner hole.
   Mid-IR interferometry offers unique constraints, complementary to mid-IR features, for studying the mineralogy in disks. A better 
   uv coverage is needed to constrain in detail the distribution of the crystalline forsterite in the disk of AC\,Her, but we
   find strong similarities with the protoplanetary disk HD100546.}


   \maketitle
%

\section{Introduction}
Post-asymptotic giant branch stars (post-AGB stars) are an evolved evolutionary phase of low- to intermediate-mass stars. They
show a large variety of circumstellar characteristics \citep{2003ARAAVanWinckel}, but a significant fraction of 
the optically bright objects show a distinctive near-IR excess \citep{2006AAdeRuyter,2014MNRASKamath}. 
This excess can be explained as being due to thermal emission of warm dust in the close environment of the central source
\citep{2006AAdeRuyter}. It is now well established that this spectral energy distribution (SED) characteristic 
indicates the presence of a stable and compact dusty
reservoir, likely a Keplerian disk \citep[e.g.][]{2006AAdeRuyter,2007AADeroo,2013AAHillen,2014AAHillen}. This was confirmed 
when the Keplerian rotation of the gas was first resolved by \citet{2005AABujarrabal} in one object, and recently endorsed 
by the most detailed position-velocity maps of the same object with the Atacama Large Millimeter 
Array (ALMA) by \citet{2013AABujarrabalC}. A survey using single-dish CO line data confirms that rotation 
is widespread \citep{2013AABujarrabalB}, which is a strong observational indicator of stability. 
While neither the formation nor the evolution of these disks is well understood, both are linked to binary
interaction processes, as evidence is mounting that sources with disk-like SEDs are indeed all binaries 
\citep{2009AAVanWinckel, 2013EASGorlova}.

The longevity of the disk is further corroborated by the strong processing of the dust grains 
as attested by the infrared spectral dust-emission features \citep[e.g.][]{2008AAGielen,2011AAGielen} and 
the mm continuum fluxes that indicate the presence of large grains \citep{2005AAdeRuyter}. 
A striking result is the large abundance and almost ubiquitous presence of crystalline grains
\citep[][]{2002AAMolsterC, 2011AAGielen}. 

Crystalline olivine grains ((Mg,Fe)$_{2}$SiO$_{4}$) are formed by high temperature ($>$1000 K) processes, 
either condensation from the gas phase or annealing of amorphous dust grains. This makes crystalline 
dust a tracer of high temperature regions and processes in the disk. In protoplanetary disks crystalline 
dust is dominantly observed in the inner parts, close to the central star where temperatures 
are high enough \citep{2004NaturVanBoekel}, but are also found farther out and are due 
to radial mixing mechanisms \citep{2004AAGail}. In situ processes that locally increase the 
temperature and pressure can also form crystalline grains at larger distances  
from the central star where it would otherwise be too cold. Examples of such mechanisms are formation 
by collisions \citep{2010IcarMorlok}, by shocks \citep{2002ApJHarker} or by stellar outbursts \citep{2009NaturAbraham}. 
Crystalline dust abundances in protoplanetary disks are also locally enhanced 
in the walls of gaps that are likely formed by planets \citep{2011AAMulders,2013AAMuldersB}. 

The spatial distribution of the crystalline dust may hold unique clues about 
the formation and/or evolution of the disks in post-AGB systems. A distribution 
equal to that of the amorphous dust might suggest that all the crystalline material was formed during the early 
formation of the disk. It is not clear whether the formation of the disk 
involved predominantly Roche-Lobe overflow or a common-envelope phase. In either case the specific 
physical and chemical conditions in various mass flows within the system may have been 
beneficial to forming crystalline dust before the disk formed. 
Conversely, a very centrally condensed distribution may be a hint that the crystalline material
formed primarily via a disk-related process. Even if radial mixing may transport some material outwards, 
the detailed spatial distribution of crystalline olivines can thus hold information about the disk's history.

One can probe the spatial distribution and grain properties of crystalline olivine by modeling its multiple mid-IR and far-IR 
spectral features. These features contain a wealth of information because their precise peak-wavelength position, relative band 
strengths, and the band shapes are dependent on the temperature, composition, and abundance of the 
grains \citep{2003AAKoike,2006MNRASSuto}.
Based on optically thin fits to the mid-IR features \citet{2008AAGielen} and \citet{2011AAGielen} found 
that the dust in post-AGB disks is relatively well mixed, with a similar dust composition throughout the disk. 
These results still need to be confirmed with full radiative transfer models to rule out any influence of 
optical depth effects. Here we present a first attempt at such a study and we show the role that mid-IR 
interferometry can play for the study of post-AGB disks, and their mineralogy.

We focus on one of the brightest systems, AC\,Her, which is also one of the most regular and best studied RV Tauri pulsators.
\citet{1998AAVanWinckel} presented the orbital solution as well as a photospheric abundance analysis, which showed the close 
resemblance between AC\,Her and the, then few, known post-AGB stars in binary systems with disks. Their detected
photospheric abundance pattern was simultaneously found by \citet{1998ApJGiridhar}.
The mm continuum fluxes have been measured by \citet{2011ApJSahai}. The Keplerian kinematics of the outer disk is resolved in 
CO rotational lines, resulting also in an estimate of the disk's mass \citep{2015arXivBujarrabal,2013AABujarrabalB}.
Several authors have attempted to image AC\,Her's disk with conflicting results. Most recently, \citet{2003ApJClose} 
and \citet{2012AAGallenne} both found AC\,Her to be unresolved in the thermal-IR on scales greater than 0.2$\arcsec$ using diffraction-limited
imaging with different instruments. A crude radiative transfer disk model was constructed by \citet{2007AAGielen}, 
who also analyzed the mid-IR features with an optically thin approach (see Sect.~\ref{section:extraobservations}).

We return to these results in the coming sections, where we discuss our observations  
(Sect.~\ref{section:extraobservations}) and our estimation of the stellar parameters (Sect.~\ref{section:stellarSED}).
The stellar properties are needed as input for our radiative transfer models (Sects.~\ref{section:MCMax},~\ref{section:structural}, 
and~\ref{section:features}). In this last section we will for the first time add crystalline olivine to a full radiative transport model and study 
the resulting features in the mid-IR, at 69 $\mu$m, and in the interferometric data.


\section{Observations} \label{section:extraobservations}
In this section we describe the sources and data reduction strategies of the data sets to be used in the remainder of 
this paper. Being an RV Tauri pulsator, the pulsation phase is taken into account in the selection of the data that are fitted.
The light curve of AC\,Her was retrieved from the AAVSO and AFOEV databases. 

\subsection{Literature photometric data} \label{subsection:photometry}
\begin{figure*}
   \centering
   \includegraphics[width=16cm,height=12cm]{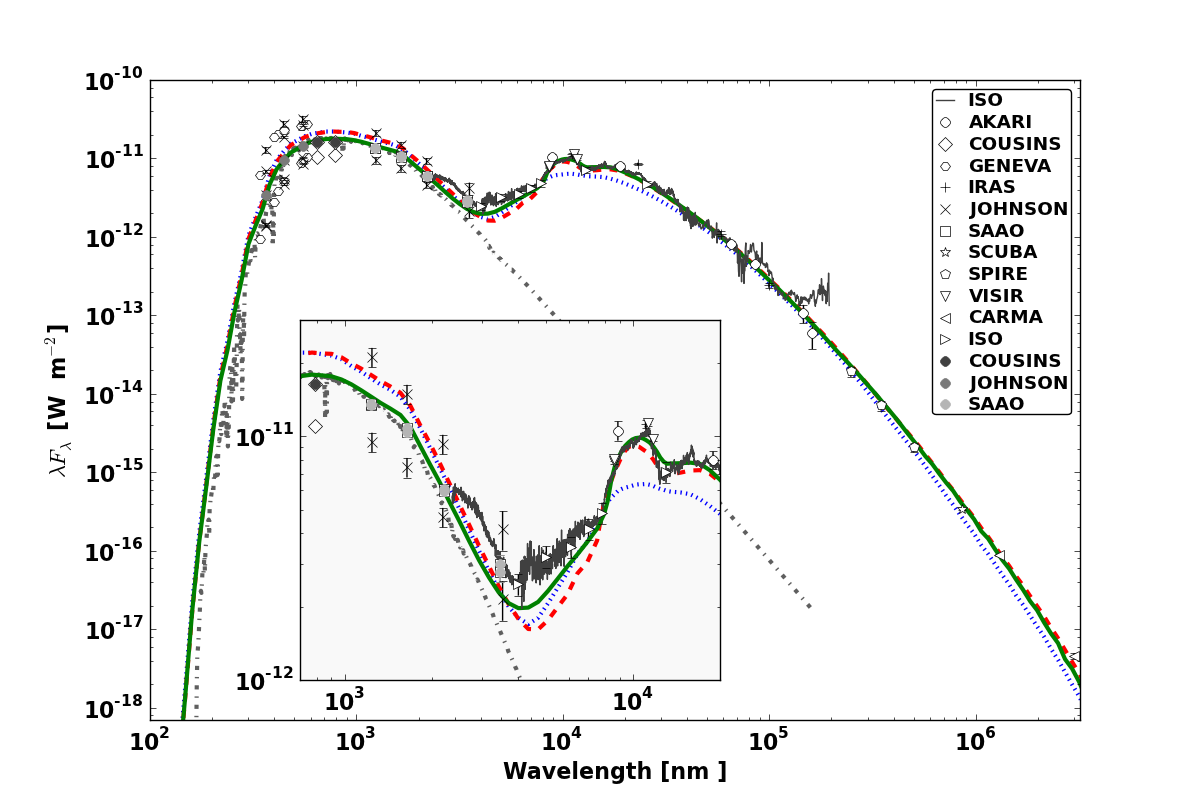}
   \caption{SED of AC\,Her. The gray dot-dashed line shows the input Kurucz model, 
     reddened with an interstellar contribution of $E(B-V)=0.135$ (see Sect.~\ref{section:stellarSED}). The dotted blue 
     line is the best single-power-law model (see Sect.~\ref{subsection:singlepowermodels}). The dashed red line 
     corresponds to our base model (i.e., the best double-power-law model obtained in Sect.~\ref{subsection:doublepowerlaws}).
     The full green line is the best-fit model with the alternative dust composition (Sect.~\ref{subsubsection:fluxdeficit}).
     The inset shows a zoom-in of the near- to mid-IR wavelength range. We note that the ISO spectrum is included in our fit through 
     the indicated anchor points (see the legend).}
     \label{figure:fullSED}
\end{figure*}

We collected photometric and spectroscopic observations of AC\,Her from various sources in the literature to construct its full spectral 
energy distribution (SED, see Fig.~\ref{figure:fullSED}).
Although AC\,Her has been observed extensively, mostly the measurements do not cover a full 
pulsation cycle. Nevertheless, we gathered optical to near-IR photometric data for three typical pulsation phases, although from 
different cycles: primary and secondary minimum as well as maximum light.

For the near-maximum and -minimum pulsation phases, we used the near-IR JHKL photometry from \citet{2010yCatTaranova} as well as the 
optical photometry in the Geneva system from \citet{2006AAdeRuyter}. In addition, we used the Johnson UBV fluxes from \citet{1979ApJSDawson} 
and \citet{1993AAZsoldos} near maximum, and from \citet{1993AAZsoldos} and \citet{1992AAShenton} near minimum. 
From this last source we also took Cousins R and I fluxes. At the secondary minimum phase, we used the Johnson-Cousins UBVRI 
data from \citet{1987MNRASGoldsmith} and the JHKL SAAO photometry from \citet{1992AAShenton}. 

At mid-IR and longer wavelengths we collected publicly available photometric data from the Infra-Red Astronomy Satellite 
\citep[IRAS,][]{1984ApJNeugebauer}, the AKARI satellite \citep{2007PASJMurakami}, as well as the Submillimetre Common-User 
Bolometer Array \citep[SCUBA][]{1999MNRASHolland} 850~$\mu$m measurement published in \citet{2006AAdeRuyter}, and the 1.3 and 3~mm fluxes measured with
the Combined Array for Research  in Millimeter-wave Astronomy (CARMA) by \citet{2011ApJSahai}. 
The fluxes extracted from VISIR images by \citet{2012AAGallenne} were also included in our analysis.

AC\,Her has been claimed to be photometrically variable even at mid-IR wavelengths \citep[][]{1992AAShenton,1971PhDTGehrz}, where 
the disk totally dominates the energy output. More recent observations of this variability are not available. 
We are confident that variability will not influence the results of this paper because 1) it is likely that only the inner rim of 
the disk will significantly respond to the varying stellar luminosity, hence fluxes beyond $\sim$25~$\mu$m should be unaffected; 2) we exclude 
the IRAS fluxes at 12 and 25~$\mu$m as well as the AKARI fluxes at 9 and 18~$\mu$m from our SED fit owing to their uncertain phase attributions; and 
3) all other data were taken near a secondary minimum phase or at phases with comparable V band magnitudes.

\subsection{SPIRE photometry}
The Spectral and Photometric Imaging Receiver \citep[SPIRE,][]{SPIREGriffin,SPIRESwinyard} instrument on board the Herschel 
satellite \citep{2010AAPilbratt} was used to acquire new far-IR photometric data of AC\,Her in February 2012 
(Herschel Observation ID 1342239790). A 169\,s exposure in the \textit{small-map photometry} observing mode was obtained simultaneously 
in three passbands, which are centered at 250\,$\mu$m, 350\,$\mu$m, and 500\,$\mu$m. We refer to \citet{2014AAHillen} for a description of 
the flux extraction procedures. The errors on the final fluxes are dominated by the absolute flux 
calibration, assumed to be at a precision level of 15\% \citep{SPIRESwinyard}. The SPIRE data are included in Fig.~\ref{figure:fullSED}.

\subsection{Spectroscopic data: ISO and PACS/Herschel} \label{subsection:ISO+PACS}
AC\,Her was observed with the short (SWS) and long (LWS) wavelength spectrometers on board the Infrared Space Observatory (ISO)
\citep{1999NaturMolster,2002AAMolster,2002AAMolsterB,2002AAMolsterC}. This spectrum was taken 
close to a secondary minimum pulsation phase. The absolute calibration of the 
spectrum is closely in agreement with the near-IR K and L bands and with the mid-IR VISIR fluxes, that all correspond to 
a similar pulsation phase, and with the IRAS and AKARI photometry beyond 25~$\mu$m, which are averages of scans taken at several epochs. 
The flux-calibrated spectrum is also included in Fig.~\ref{figure:fullSED}.

The full spectrum shows prominent spectral bands in the 
mid-IR from crystalline olivine grains \citep[][]{1999NaturMolster}, mainly from its Mg-rich end-member forsterite. Typical 
features of magnesium-rich crystalline olivine are located at 11.3, 16.2, 19.7, 23.7, 28.0, and 33.6~$\mu$m. The most recent analysis of 
these features was performed by \citet{2007AAGielen}. They fitted the mid-IR spectral bands 
with an optically thin model. Their best fit included a hot (800 K) population and an almost as 
abundant cold (100 K) population of forsterite grains. Both populations contained small (0.1 $\mu$m) and large 
(1.5 $\mu$m) forsterite crystals. 

Crystalline olivine has now also been detected in the far-IR at 69~$\mu$m. This far-IR spectral band is a 
strong diagnostic of especially the crystal composition and the crystal temperature \citep{2003AAKoike, 2006MNRASSuto}, 
very complementary to the mid-IR features. The 69~$\mu$m band shifts to the red and broadens as a 
function of increasing grain temperature and iron content. The Photodetector Array Camera and Spectrometer (PACS) 
on board the Herschel Space Observatory \citep{2010AAPilbratt,2010AAPoglitsch}
was used to take a spectrum of AC\,Her in the wavelength range of 67-72~$\mu$m. 
The observation was performed as part of the GT program ``Forsterite dust in the circumstellar environment of evolved stars'' 
(\detokenize{GT1_jblommae_1}). The spectrum has been published by \citet{2014AABlommaert} as part of a large and 
diverse sample of evolved stars targeted with PACS. Details about the adopted observational setup and the data 
reduction procedure, can be found in the same paper.

Analysis of the peak wavelength positions and shapes of all these spectral bands show that the crystalline olivine 
grains contain no iron and are purely forsteritic \citep[Mg$_{2}$SiO$_{4}$,][]{1999NaturMolster, 2014AABlommaert}.
\citet{2014AABlommaert} moreover showed that the 69 $\mu$m band is dominated by flux from $\sim$200~K forsterite 
grains and they found a peak-to-continuum-strength ratio of the 69~$\mu$m band of 0.06 $\pm$/-0.004. 

\subsection{Mid-IR interferometry} \label{subsection:MIDIdata}
We used the MID-infrared Interferometric instrument (MIDI) on the Very Large Telescope Interferometer (VLTI) to observe AC\,Her with the 
Unit Telescopes (UTs) during three nights in 2005. On each night a single MIDI observation was acquired in the GRISM spectral mode (R=230) 
with the UT3-UT4 baseline on 25 May (baseline length and position angle: BL$\sim$60~m, PA$\sim$104$^\circ$), and 
with the UT2-UT3 baseline on 26 June (BL$\sim$45~m, PA$\sim$47$^\circ$) and on 23 July 
(BL$\sim$37~m, PA$\sim$50$^\circ$). We reduced the data with version 2.0 of the Expert Work Station (EWS) software 
\citep{2004SPIEJaffe}\footnote{\texttt{http://home.strw.leidenuniv.nl/$\sim$jaffe/ews/index.html}}.

Calibration of the raw target visibility occurs by division with the instrumental visibility ($V_\mathrm{ins}$, i.e., the interferometer's response 
to a point source) for each wavelength. The $V_\mathrm{ins}$ is interpolated from calibrator observations for which the intrinsic visibility 
is known and that were obtained immediately before and after the science measurement. Variability of the transfer function is taken into account by 
adding to the errors in quadrature the standard deviation of the $V_\mathrm{ins}$, including all calibrators within four hours of the 
science measurement. The resulting errors are very small, which attests to the good quality of all three nights.

The three MIDI observations are separated in time by about one month, corresponding to pulsation phases of $\sim$0.8, $\sim$0.3, and $\sim$0.55, 
respectively, so we checked whether the data show any sign of variability. In the V passband, the corresponding fluxes differ 
by at most half a magnitude. Moreover, direct calibration of MIDI's total flux spectrum for each of the three observations reveals 
no detectable difference, except that all three are consistently above the ISO fluxes by $\sim$20\%. Finally, even if the total flux is 
intrinsically variable in the AC\,Her system, the visibilities are likely to remain unaffected, because the physical morphology 
of the object cannot change on such short timescales. We conclude that variability is not an issue for these MIDI data.

In addition to the visibility amplitudes, MIDI measures differential phases. These are not the real Fourier phases, but 
have a linear component in wavenumber (i.e., inverse wavelength) subtracted \citep{2004SPIEJaffe}. The raw differential phases are corrupted 
by the different amount of water vapour that is experienced by the two beams, as a result of their different path lengths in the 
delay-line system. We calibrate the science data by subtracting the introduced phase shift, which is estimated by 
linearly interpolating the raw calibrator differential phase versus delay-line path length difference (per wavelength bin). 
The error on the differential phase is taken as the standard deviation of the calibrator measurements around this 
linear relation, added in quadrature to the statistical error on the phase. Differential phases can be used as a general 
indicator for asymmetry of the observed object, but they can also be included directly in model fitting. We refer 
to, e.g., \citet{2014AATristram} or \citet{2007AADerooB}, for some successful examples of differential-phase modeling and for a 
deeper discussion on its diagnostic power. In summary, one can state that the differential phase measures the wavelength 
dependence of the symmetry of the object: if the object has different asymmetries across wavelengths, the differential 
phase will be nonzero.

\section{The central star(s)} \label{section:stellarSED}
To construct a proper radiative transfer model of the circumbinary disk in the AC\,Her system 
requires correctly configuring the properties of the central heating source, i.e., defining the fundamental parameters 
of the post-AGB star. 

\begin{table}
\caption{Stellar and orbital parameters of AC\,Her.}             
 \label{table:parameters}      
 \centering                          
 \begin{tabular}{l c c c c}        
 \hline\hline                 
   & Parameter & Value & Error & Source \\    
 \hline
 \hline                        
  Literature  & Orb. Period (d) & 1194 & 6 & 1 \\
              & a$_1$ sin i (AU) & 1.39 & - & 1 \\
              & e & 0.12 & 0.02 & 1 \\
              & Mass function (M$_\odot$) & 0.25 & - & 1 \\
              & Pulsation Period (d) & 75.0 & - & 2 \\
  Fitted & T$_{\mathrm{eff}}$ (K) & 5225 & 125 & - \\
         & $\theta_\star$ (mas) & 0.353 & 0.012 & - \\
         & $\log g$ & 0.65 & 0.20 & - \\
         & E$_{\mathrm{ISM}}$(B-V) (mag) & 0.14 & Fixed & -\\
         & [Fe/H] (dex) & -1.5 & Fixed & - \\
         & i\tablefootmark{a} ($^\circ$) & 50 & 8 & - \\
  Derived & d\tablefootmark{b} (kpc) & 1.6 & 0.3 & - \\
          & L\tablefootmark{b} (L$_\odot$) & 2500 & 900 & - \\ 
          & R\tablefootmark{b} (R$_\odot$) & 61 & 12 & - \\
          & M$_1$\tablefootmark{c} (M$_\odot$) & 0.6 & 0.2 & - \\    
          & M$_2$\tablefootmark{d} (M$_\odot$) & 1.2 & 0.2 & - \\
          & a (AU) & 2.7 & 0.2 & -\\

 \hline
 \end{tabular}
 \tablefoot{\tablefoottext{a}{From the interferometric fit, see Sect~\ref{section:structural}.}
            \tablefoottext{b}{From the PL-relation of \citet{1998AJAlcock}.}
            \tablefoottext{c}{From the fitted $\log g$ and derived radius.}
            \tablefoottext{d}{From M$_1$, the mass function and the fitted inclination.}}
 \tablebib{(1) \citet{1998AAVanWinckel}, (2) \citet{2009yCatSamus}}
 \end{table}

Table~\ref{table:parameters} gives an overview of the adopted values for the relevant systemic 
and stellar parameters. There are three groups of parameters: those that have no direct impact on any of our observables, i.e., the orbital elements
and pulsation period, taken from \citet{1998AAVanWinckel} and from the General Catalogue of 
Variable Stars \citep[GCVS,][]{2009yCatSamus}, respectively; parameters that do have an influence and are therefore fitted; and 
parameters that are indirectly derived from the former groups.

The parameters that affect the SED are the effective temperature T$_{eff}$, the gravity $\log g$, 
the angular diameter $\theta_\star$, the interstellar reddening E$_{\mathrm{ISM}}$(B-V), and the metallicity [Fe/H]. The metallicity was determined
spectroscopically by \citet{1998AAVanWinckel} and by \citet{1998ApJGiridhar}. The values derived by these authors are  
consistent within the errors, but we assume the value derived by \citet{1998AAVanWinckel} since their T$_{eff}$ and $\log g$ are in better 
agreement with what we find in the following. We also fix the value of E$_{\mathrm{ISM}}$(B-V) based on the consistent 
values of E$_{\mathrm{ISM}}$(B-V)$\sim$0.14 derived from the Galactic extinction maps of \citet{1998ApJSchlegel} and \citet{2001ApJDrimmel}.

The T$_{eff}$, $\log g$, and $\theta_\star$ are subsequently fitted to the stellar part of the SED, using the grid-based method detailed 
in \citet{2011AADegroote}. With this approach correlations between parameters can be easily identified and 
taken into account in determining the final uncertainties. The goodness of fit, and the confidence intervals (CI) derived from it, are determined 
with a $\chi^2$ statistic; in this case with three degrees of freedom. In practice, a Kurucz model SED is first reddened with
the reddening law of \citet{2004ASPCFitzpatrick}, subsequently integrated over the relevant photometric passbands, and ultimately scaled to 
the measured fluxes by optimizing the angular diameter. The resulting best-fit parameter values are listed in Table~\ref{table:parameters} and the 
CI are shown in the online Figs.~\ref{figure:CIteff} and~\ref{figure:CItheta}.

\onlfig{
\begin{figure}
   \centering
   \includegraphics[width=8cm]{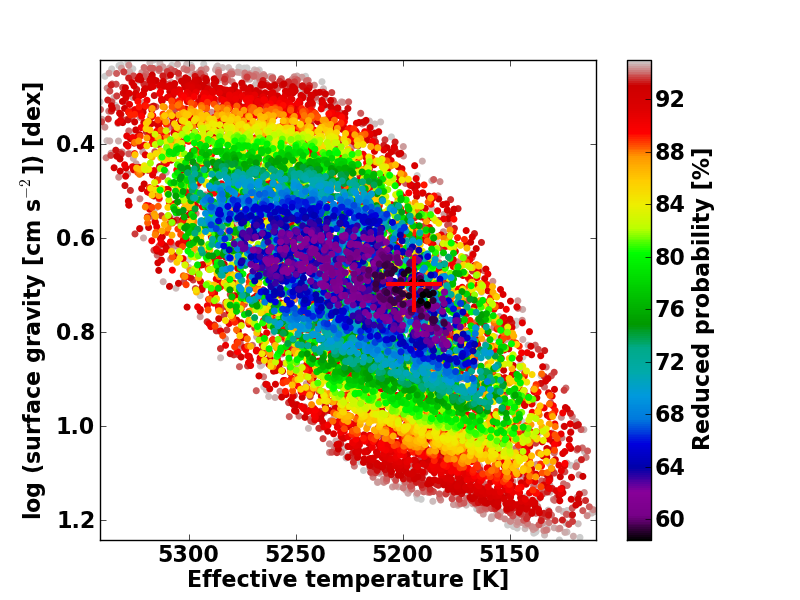}
   \caption{Confidence interval of the stellar SED fit: T$_{eff}$ versus $\log g$.
    }
    \label{figure:CIteff}
\end{figure} 
}

\onlfig{
\begin{figure}
   \centering
   \includegraphics[width=8cm]{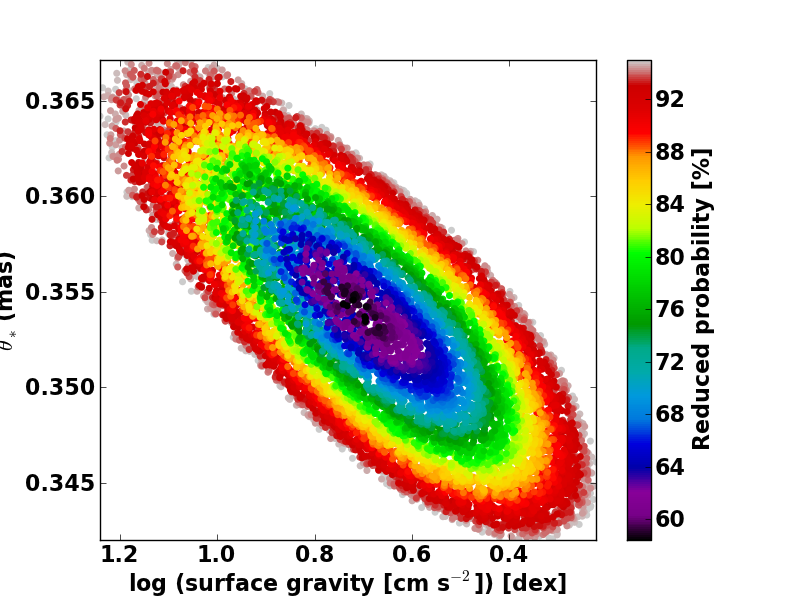}
   \caption{Confidence interval of the stellar SED fit: $\log g$ versus $\theta_\star$.
    }
    \label{figure:CItheta}
\end{figure} 
}

\begin{figure}
   \centering
   \includegraphics[width=\hsize]{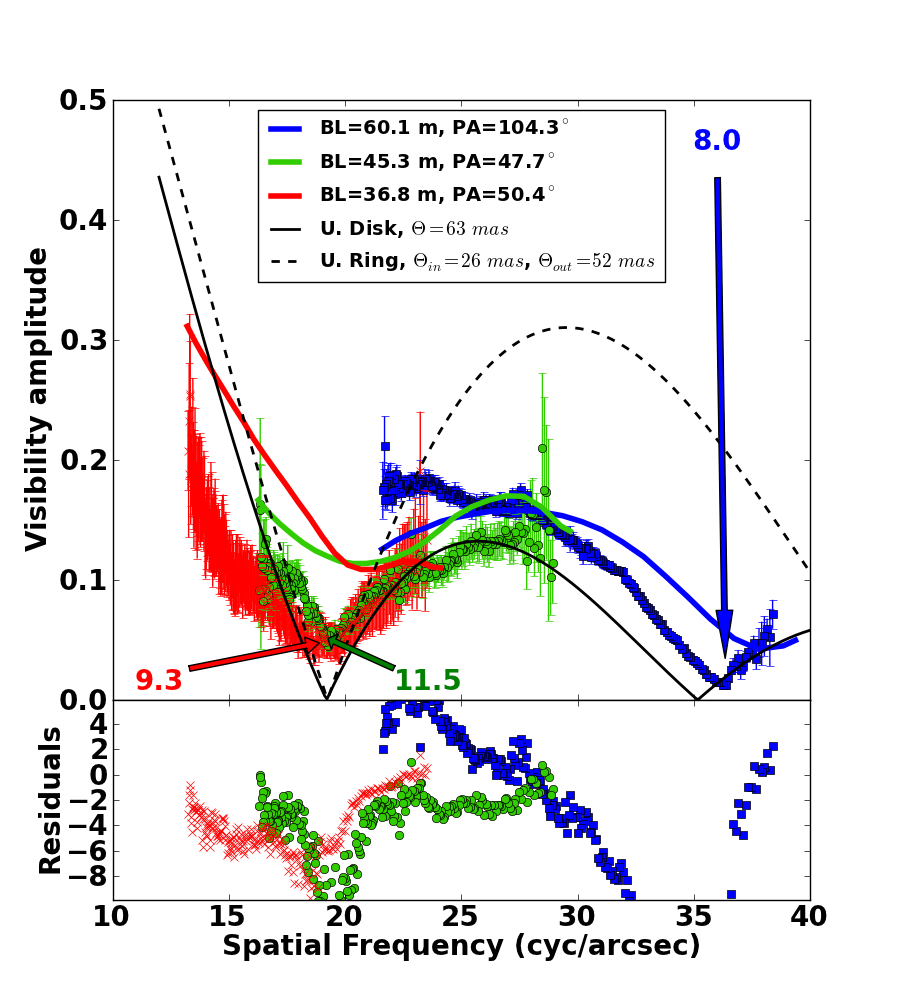}
   \caption{Upper panel: the measured MIDI visibilities of AC\,Her, color coded with respect to the baseline as listed in the legend, 
   overplotted (full thick lines) with the best-fit single-power-law radiative transfer model (see Sect.~\ref{section:structural}), as a 
   function of spatial frequency. Also shown in black are the uniform disk and uniform ring models mentioned in Sect.~\ref{section:simpleAnalysis}.
   Lower panel: the residuals with respect to the radiative transfer model. 
     }
     \label{figure:MIDIvis-sf}
\end{figure} 

\begin{figure}
   \centering
   \includegraphics[width=\hsize]{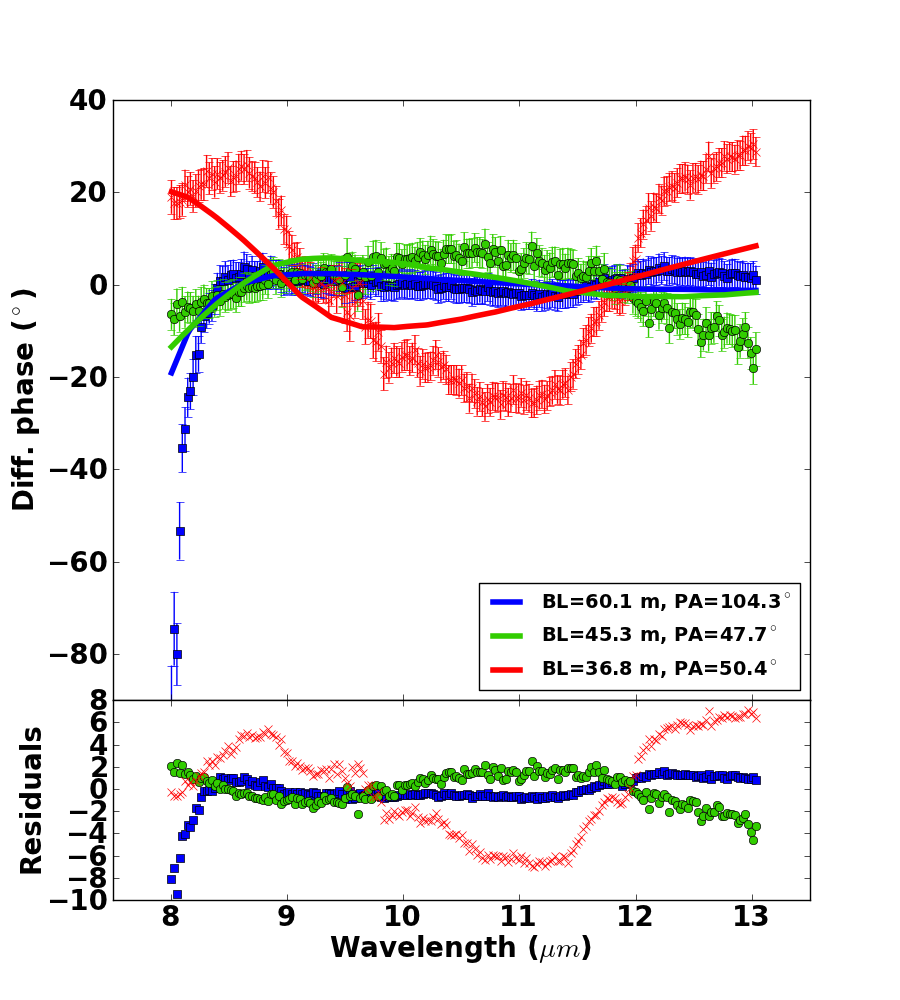}
   \caption{Upper panel: the MIDI differential phases of AC\,Her, similarly color coded and overplotted with the same radiative transfer model 
   as in Fig.~\ref{figure:MIDIvis-sf}, as a function of wavelength. Lower panel: the residuals with respect to the radiative transfer model.}
   \label{figure:MIDIDP-wl_res}
\end{figure} 

We use the P-L relation of \citet{1998AJAlcock} to estimate the absolute V band magnitude that corresponds to the pulsation 
period of AC\,Her. We integrate the best-fit SED model over the V passband and thus find a distance of 1.6$\pm$0.3~kpc for a corresponding
luminosity of 2500~L$_\odot$.

Finally, we note that any scattering contribution to the stellar part of the SED is not properly taken into account in this analysis. 
\citet{2013AAHillen} recently found --on the basis of a combined 
optical and near-IR interferometric data set-- that a significant fraction ($\sim$35\%) of all optical light in a similar system, 89 Herculis, 
is in fact circumstellar scattered light. For no other post-AGB binary system has the scattered light fraction been measured yet.
We assume that the scatter in the P-L relation of \citet{1998AJAlcock} incorporates a similar spread in 
scattered light fraction as is present in the Galactic RV Tauri population, so that population-wise the estimated distances should 
be correct. For any individual system, however, the stellar luminosity is likely biased. Since we do not have data 
to constrain the scattered light fraction of AC\,Her, we ignore the optical to near-IR part of 
the SED for the rest of this paper.


\section{A simple analysis of the interferometric data} \label{section:simpleAnalysis}
Before continuing to the radiative transfer analysis of our data, we analyze the interferometric data on a qualitative level.

\begin{figure}
   \centering
   \includegraphics[width=8cm,height=6cm]{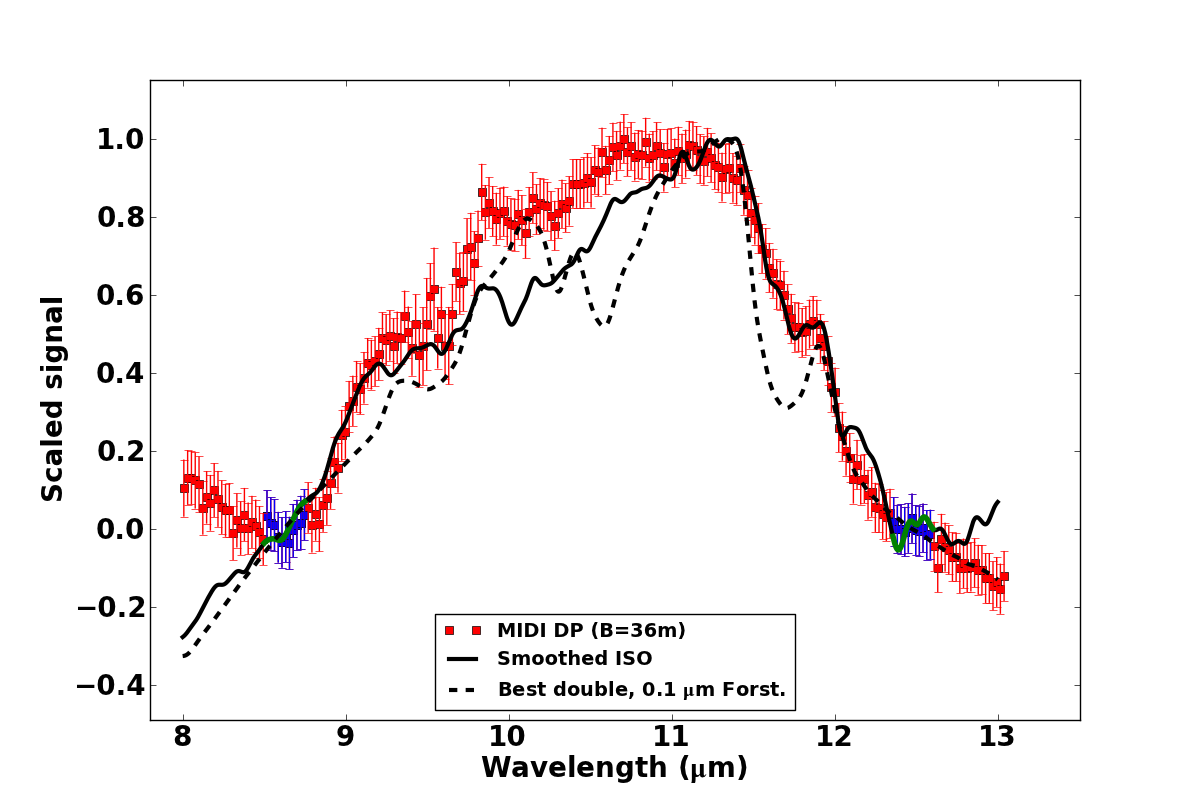} 
   \caption{Comparison between the differential phase signal on the shortest MIDI baseline and the ISO spectrum in the same wavelength range. 
   The red squares show the MIDI differential phases after subtraction of a linear trend (fit to the blue points), inversion, and 
   normalization. The ISO spectrum was first smoothed with a Gaussian kernel ($\sigma = 0.025 \mu$m, comparable to the resolution of the MIDI grism 
   at 13~$\mu$m). The full black line shows this smoothed ISO spectrum after subtraction of a linear trend (fit to the green points) and normalization. 
   The differential phase of the intermediate baseline can be scaled similarly, but without sign inversion and with a smaller normalization factor, 
   because these spatial frequencies fall largely in the second lobe of the radial visibility curve along this direction. The 
   ISO spectrum of the base model added with 1\% of 0.1~$\mu$m forsterite grains (see Sect.~\ref{section:features}) 
   is shown as the dashed line. We note that this model is normalized independently.}
     \label{figure:MIDIDPcomparison}
\end{figure}

The calibrated MIDI visibilities are shown in Fig.~\ref{figure:MIDIvis-sf} as a function of spatial frequency. 
The differential phases are shown in Fig.~\ref{figure:MIDIDP-wl_res} as a function of wavelength. Thanks to the spectral range
of MIDI that encloses the N band (8-13~$\mu$m), and the limited range in baseline length, the three observations overlap in their radial 
spatial frequency coverage. From the shapes of the three visibility and differential phase spectra, we can already draw several conclusions 
independent of any quantitative modeling.
\begin{itemize}
 \item The two shortest-baseline observations, which have similar position angles, are consistent with each other and show a clear minimum in 
 visibility amplitude at more or less the same radial spatial frequency, hence at different wavelengths (see Fig.~\ref{figure:MIDIvis-sf}). 
 This implies that the morphology of the source, and not its spectral characteristics, is controlling the shape of the visibility spectrum 
 at these spatial frequencies. 
 \item The measured visibility amplitudes are overall very low, i.e., the data probe high spatial frequencies near two 
 consecutive visibility minima. The ``global geometry'' of the system thus dominates the shape of the observed
 visibility spectra.
 \item The visibility amplitude minima do not reach a full null. For the 60~m baseline (blue squares), the minimum amplitude 
 is reached near 8~$\mu$m and is $\sim$0.012-0.025. 
 This is consistent with the estimated fraction of unresolved stellar flux of $\sim$2\%, based on the directly calibrated total MIDI 
 flux and the SED model found in Sect.~\ref{section:stellarSED}. This means that the circumstellar component does pass through a null at 
 this baseline. The minimum visibilities at the shorter baselines are reached near 9.3 and 11.5~$\mu$m, respectively, and 
 are in the range 0.04-0.06, significantly above the estimated stellar flux fractions of $\sim$1\% and $\sim$0.5\% at 
 the respective wavelengths.
 \item A strong differential phase signal (see Fig.~\ref{figure:MIDIDP-wl_res}) is detected on the 60~m baseline, and 
 moderate signals are present on the 45 and 36~m  baselines (green dots and red crosses, respectively). 
 The differential phase on the 60~m baseline seems to be dominated by the source 
 morphology. The strong signal near 8~$\mu$m is a clear example of a 180$^\circ$ phase jump (which is the result of a crossing of a visibility 
 null); the signal is a result of the changing spatial resolution of the interferometer with wavelength (see the previous item). 
 The fact that this transition is not step-like, but smooth yet steep, implies that the source morphology along this baseline 
 direction is not entirely point-symmetric but is close to it. The shape of the differential phase signal at the shortest baselines is remarkably 
 similar to the silicate feature in the ISO spectrum, as is illustrated in Fig.~\ref{figure:MIDIDPcomparison}. The fact that the 
 feature induces a signal in the differential phase at these baselines means that the position of the photocenter of the emission folded 
 along this baseline direction is different between the continuum and the feature.
\end{itemize}

To define a range of sensible inner rim radii for our radiative transfer disk models, we first estimate the overall size of the mid-IR emission
with simple geometric models. Figure~\ref{figure:MIDIvis-sf} shows two such models: a uniform disk (UD) and a
uniform ring (UR), as the full and dashed black lines, respectively. The UD diameter and UR inner diameter have been optimized so that the curves
reach their first null coincident with the observed position on the two shortest baselines. In this exercise we keep the UR's 
outer-to-inner-diameter ratio fixed at two. We find that the circumstellar emission seems to arise from within 
$\sim$50-60~mas from the central source. Of course, these simple models cannot account for the full complexity of the wavelength-dependent 
source morphology, which is influenced by optical depth effects and by the thermal structure of the emitting material. 
Nevertheless, based on the visibility amplitudes, we can conclude that the source has a brightness distribution that 
does not look too different from a UD overall. 

A circularly symmetric brightness distribution cannot, however, explain the striking features listed above: it does not fit with the 
shifted photocenter in and outside the silicate feature, nor with the smooth differential phase signal on the longest baseline or with the higher 
visibility plateau in the first null. 

In the following sections we show that a circumbinary disk can explain the MIDI data set naturally, as well as the SED.

\section{Radiative transfer modeling} \label{section:MCMax}
Radiative transfer disk models of AC\,Her have been computed previously by \citet{2007AAGielen} with the 2D Monte Carlo
radiative transfer code MCMax. Here we use the code with different settings and include additional physics. The models of \citet{2007AAGielen}
were not yet computed with a grain size distribution up to mm-sized grains in combination with a full treatment of dust settling, which are both 
needed to fit the far-IR to the mm part of the SED. In all the models presented here, these effects are included.

\subsection{The code} \label{subsection:code}
The MCMax code \citep{2009AAMin} has been widely and successfully applied to model observables obtained for gas-rich disks in a variety of 
environments \citep{2011AAMulders,2011AAVerhoeff,2013AAAcke}. We refer to these papers for a full description of 
the basic features of the MCMax code. To include the settling of dust we follow the approach that was presented in \citet{2012AAMulders}, and 
first applied in a post-AGB context in \citet{2014AAHillen}.
In short, MCMax first computes the radiative equilibrium temperature structure throughout the disk. The vertical density structure 
of the disk is then self-consistently determined by solving the equations of vertical hydrostatic equilibrium, 
implicitly assuming thermal coupling between the gas and dust, but taking dynamical decoupling to model dust settling. Both 
steps are iterated until convergence is reached. Dust scattering can be treated in a full angle-dependent way
or, to speed up the computations \citep{2009AAMin}, by assuming isotropic scattering. We apply isotropic scattering in this paper.

\subsection{The dust model(s)} \label{subsection:dust}
Two different compositions are tried for the amorphous dust component in this paper. Our standard mixture consists of the composition 
measured in the interstellar dust towards the Galactic center by \citet{2007AAMin} --12.11\% MgFeSiO$_4$, 33.63\% Mg$_2$SiO$_4$, 37.67\% MgSiO$_3$,
1.58\% NaAlSi$_2$O$_6$-- co-added with 15\% metallic iron (all mass fractions). The shapes of the grains are 
irregular, approximated by the distribution of hollow spheres \citep[DHS,][]{2005AAMin}.
Next to the standard composition, we also computed a grid of models consisting of only MgFeSiO$_4$ 
(see Sect.~\ref{subsection:doublepowerlaws}) the opacities of which were computed from the optical 
constants of \citet{1995AADorschner} in the same way as before (f$_{\rm{max}} = 0.8$). 

The forsterite opacities that will be added to the model in Sect.~\ref{section:features} are calculated from the optical constants 
of \citet{2006MNRASSuto} in the same manner as the amorphous grains (using DHS). Since the forsterite opacities are temperature 
dependent, we use optical constants at the temperatures of 50, 100, 150, 200, and 295~K. 
MCMax can use temperature-dependent opacities \citep[for more details, see, e.g.,][]{2014AADeVries,2011AAMulders}, meaning the model 
correctly predicts the temperature-dependent spectral features of the forsterite grains. 


\begin{table*}
 \caption{Parameters of the MCMax radiative transfer disk models. Our adopted base model is highlighted in boldface.}             
 \label{table:MCMaxmodels}      
 \centering                          
 \begin{tabular}{l l c c c c}        
 \hline\hline                 
   & Parameter & Value(s) & Best Single & \textbf{Best Double} & Best Alt. \\    
 \hline
 \hline                        
  Fitted & M$_{\mathrm{dust}}$ & 10$^{-4}$-10$^{-2}$(x2) M$_\odot$ & 1\,10$^{-3}$ & \textbf{2.5\,10$^{-3}$} & 2.5\,10$^{-3}$  \\
         & gas/dust & 1.0;10.0;100.0 & 100 & \textbf{10} & 10  \\
         & q & 3.00;3.25;[3.35]\tablefootmark{a} & 3.00 & \textbf{3.25} & 3.00  \\
         & p$_{\mathrm{in}}$ & -0.5;-1.0;-2.0;-3.0;\{-4.0;-5.0;-6.0;-7.0\}\tablefootmark{b} & - & \textbf{-3.0} & -3.0 \\
         & a$_{\mathrm{max}}$ & 0.3;1.0 mm\tablefootmark{c} & 1.0 & \textbf{1.0} & 1.0 \\
         & R$_{\mathrm{in}}$ & 16-44(+2)~AU & 36 & \textbf{34} & 32  \\
         & R$_{\mathrm{mid}}$/R$_{\mathrm{in}}$ & 1.25;1.50;2.00;2.50 & - & \textbf{2.0} & 2.0 \\
         & i & 10-80$^\circ$ & 40 & \textbf{50} & 55 \\
         & PA & 0-360$^\circ$ E of N & 305 & \textbf{300} & 300 \\  
  Fixed  & p$_{\mathrm{out}}$ & 1.0 & & & \\
         & a$_{\mathrm{min}}$ & 0.01~$\mu$m & & & \\
         & R$_{\mathrm{out}}$ & 200 AU & & & \\
         & $\alpha$ & 0.01 & & & \\
         & comp. & standard/alternative & standard & \textbf{standard} & alternative \\         
  $\chi^2$(/min) & SED & - & 18 (3.3) & \textbf{8.8 (2.9)} & 4.5 (1.7) \\
           & VIS & - & 38 (1.7) & \textbf{6.5 (2.3)} & 4.7 (1.1) \\
           & DP & - & 9.5 (1.6) & \textbf{5.5 (1.4)} & 6.2 (1.5) \\
 \hline
 \end{tabular}
 \tablefoot{\tablefoottext{a}{The values between brackets are only included in the single power-law grid.}
            \tablefoottext{b}{The values between curly braces are only computed in the extended grid (see Sect.~\ref{subsubsection:extendedgrid}).}
            \tablefoottext{c}{Our data are not sensitive to larger grain sizes, which, moreover, only increase the total dust mass further.}}
            
\end{table*}

\section{A structural model} \label{section:structural}
 \subsection{The strategy}
 Extensive grids of radiative transfer disk models are to be compared to our 
 observations: the SED, the MIDI visibilities and the MIDI differential phases. The goal is to infer the density and thermal 
 structure of the circumstellar material.
 
 Although MCMax self-consistently computes the vertical structure of the disk, the radial (or surface density) distribution 
 of the material is required as input for the models. Similarly to the disk in the 89 Her system \citep{2014AAHillen}, we start by 
 parameterizing the surface density distribution with a single power law, $\Sigma \propto r^{-p_{\mathrm{out}}}$. 
 
 Throughout this paper we keep the turbulent mixing strength $\alpha$ fixed at 0.01 \citep{1998ApJHartmann}, the outer disk 
 radius at 200~AU, and the minimum dust grain size at 0.01~$\mu$m, since these parameters either have a minimal influence on 
 our observables or are coupled with other parameters (e.g., $\alpha$ and the gas/dust ratio). The value of 
 $p_{\mathrm{out}}$, which can only be constrained with resolved millimeter observations, is fixed at 1.0, the typical 
 value found in protoplanetary disks \citep[][]{2011ARAAWilliams}.

 The remaining free physical parameters are the inner radius R$_{\mathrm{in}}$, the global gas/dust ratio, the grain size 
 distribution power-law index q, the maximum grain size, and the total dust mass M$_d$. 
 In addition, there are two geometrical parameters: the inclination $i$ and the position angle of the disk on the sky.
 Table~\ref{table:MCMaxmodels} lists the allowed values for each parameter in our grid.
 
 For each model, the computed spectrum is reddened with E$_{\mathrm{ISM}}$(B-V)=0.14, as determined in Sect.~\ref{section:stellarSED}, and integrated
 over the photometric passbands. We include the ISO spectrum by taking the monochromatic flux 
 at seven continuum wavelengths in the 4-25~$\mu$m region (see the legend in Fig.~\ref{figure:fullSED}).
 MCMax provides the complex visibility for each model, from which we extract the visibility amplitude and compute the 
 MIDI differential phase. To save computation time, we calculate complex visibilities only for 20 wavelengths between 7.5 
 and 13.5~$\mu$m, which are then linearly interpolated to the observed wavelength grid when calculating synthetic 
 visibilities and differential phases. For each observable a separate $\chi^2$ is computed, i.e., for the SED, the MIDI visibility 
 amplitudes and the MIDI differential phases. 
  
 To find models that fit all the data sets, we use the following criterion.
 A model is selected as good enough if it satisfies $\chi^2 < 4 \times \max(1.0,\chi^2_{\mathrm{min}})$ for \textit{all} of the data sets.
 The lower limit of 1.0 is imposed to the $\chi^2_{\mathrm{min}}$ to avoid excluding too many models if a particular data set is 
 over-fitted, which we do not expect given the complexity of these objects. 
  
 \subsection{Single-power-law models} \label{subsection:singlepowermodels}
 Although an extensive grid of models was computed, only two of our single-power-law models conform to the acceptance criterion. 
 The best of these two is included in Table~\ref{table:MCMaxmodels} (``Best Single'') and 
 Figs.~\ref{figure:fullSED},~\ref{figure:MIDIvis-sf}, and~\ref{figure:MIDIDP-wl_res}.
 The $\chi^2$ values for both models are still high, $\chi^2_{\rm{SED}} \geq 18$, $\chi^2_{\rm{VIS}} \geq 38$, and $\chi^2_{\rm{DP}} \geq 9.5$. 
 In particular the visibilities and the SED do not fit well. The differential phases are already well represented by both 
 models. Yet the amplitude at the shortest baseline is smaller than observed, because 
 the silicate feature is almost completely absent in this model owing to a lack of small grains (see also the SED 
 in Fig.~\ref{figure:fullSED}). The SED does not fit very well in the 10-50~$\mu$m 
 nor in the mm regime. It is also apparent that the shape of the visibility amplitude spectra 
 agree well, but the absolute scaling for the shortest baselines do not. 
 In summary, we interpret the general good agreement, i.e., the visibility 
 and differential phase shapes, as a confirmation that a well-oriented disk model captures the basic features of the source morphology.
 The orientation of the disk on the sky is strongly constrained by the combined observations at different baseline 
 position angles, and in particular by the long baseline that probes a secondary null and 
 is almost aligned with the major axis of the disk. The remaining discrepancies can be attributed to the detailed 
 inner rim shape. Because of the single-power-law parameterization of the surface density, the inner rim of these models is wall-like 
 \citep[see also][]{2014AAHillen,2010ARAADullemond}. This explains the lower inclination compared to 
 the best-fit models in the following sections: when the inclination is increased to 50$^\circ$, the $\chi^2_{VIS}$ increases significantly
 because the inner rim appears too sharp.
 
%

 \subsection{Double-power-law models} \label{subsection:doublepowerlaws}
 \subsubsection{Our standard model} \label{subsubsection:standardmodel}
 To improve upon our fit, we explore double-power-law models that smooth the inner disk rim.
 We parameterize the radial surface density profile by two power laws that are joined at an intermediate 
 radius R$_{\mathrm{mid}}$. This introduces two new 
 parameters: the R$_{\mathrm{mid}}$/R$_{\mathrm{in}}$ ratio and p$_{\mathrm{in}}$. We therefore fix the value of the 
 grain size distribution power-law index at 3.25, given the relative lack of small grains compared to the observations in our best-fit 
 single-power-law model. We allow p$_{\mathrm{in}}$ to be between -0.5 and -3.0, 
 while the turnover radius can be up to 2.5 times the inner radius.
 
\begin{figure}
   \centering
   \includegraphics[width=8cm,height=6cm]{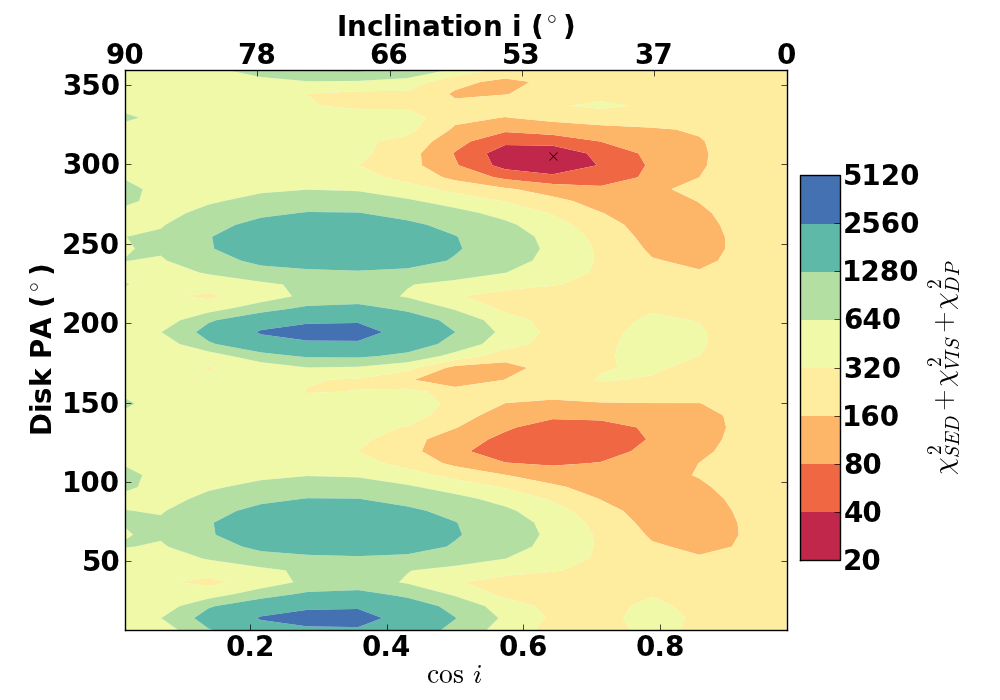}
   \caption{$\chi^2$ map of the disk inclination versus its orientation on the sky. We note that we adopt the convention that a 
   PA of 0$^\circ$ corresponds to a disk oriented with its major axis along the N-S direction and with its far (prominent) side directed
   east.}
     \label{figure:inclination-PA}
 \end{figure}

 Four models satisfy our acceptance criterion, of which two fit slightly better. 
 We include the one with the smallest dust mass (see Sect.~\ref{subsection:abundances}) in Table~\ref{table:MCMaxmodels} 
 as ``Best Double''. The values of $\chi^2$ have significantly improved: $\chi^2_{\rm{SED}} \geq 8.8$, $\chi^2_{\rm{VIS}} \geq 6.5$ and 
 $\chi^2_{\rm{DP}} \geq 5.5$. This model is used as our base model in 
 the remainder of this paper. Assuming this base model for the disk structure, the distribution of allowed 
 inclinations and PAs is illustrated by the $\chi^2$ map in Fig.~\ref{figure:inclination-PA}. There is a clearly 
 defined minimum at an inclination and PA of 50$^\circ$ and 305$^\circ$, respectively. Another minimum occurs at the same inclination 
 but at a PA of 125$^\circ$, because only the differential phases allows the 180$^\circ$ ambiguity in the disk 
 position angle on the sky to be resolved. All well-fitting models have inclinations and PAs that fall within the first contour. 
 
 \onlfig{
 \begin{figure}
   \centering
   \includegraphics[width=8cm]{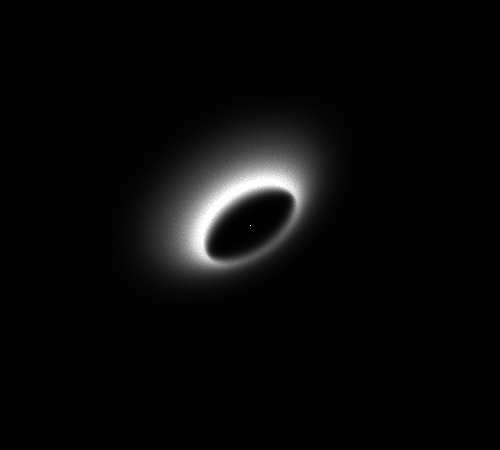} 
   \caption{A visual rendering of our best-fit double power-law model at a wavelength of 12.5~$\mu$m. The intensity is shown on a linear scale.
   North is up and East is to the left. The field-of-view of this image is 200 by 200~mas.}
     \label{figure:image}
 \end{figure} 
 }

 Although each of the model parameters 
 influences our observables (or at least some of then), we do not expect to accurately constrain all parameters independently. This is consistent 
 with the finding that a set of models fits our data equally well, so some degeneracy remains (see also Sect.~\ref{subsubsection:extendedgrid}). 
 For certain parameters there are, however, clear trends. The inner radius is well constrained
 by our interferometric data at 34$^{+8}_{-2}$~AU. The dust mass is also constrained to be 2.5-5.0$\cdot 10^{-3}$~M$_\odot$.
 The gas/dust ratio, which influences the data by its effect on the inner rim shape and on the feature/continuum ratio, is also 
 smaller than 100 in all these models. Although this effect correlates with the turbulent mixing strength parameter $\alpha$, we
 note that there are other indications for a low gas mass (see Sect.~\ref{subsection:abundances}).
 The preferred ratio R$_{\mathrm{mid}}$/R$_{\mathrm{in}}$ is clearly 2.0 or 2.5, and mostly p$_{\mathrm{in}}$ is -3.0. 
 
\begin{figure}
   \centering
   \includegraphics[width=\hsize]{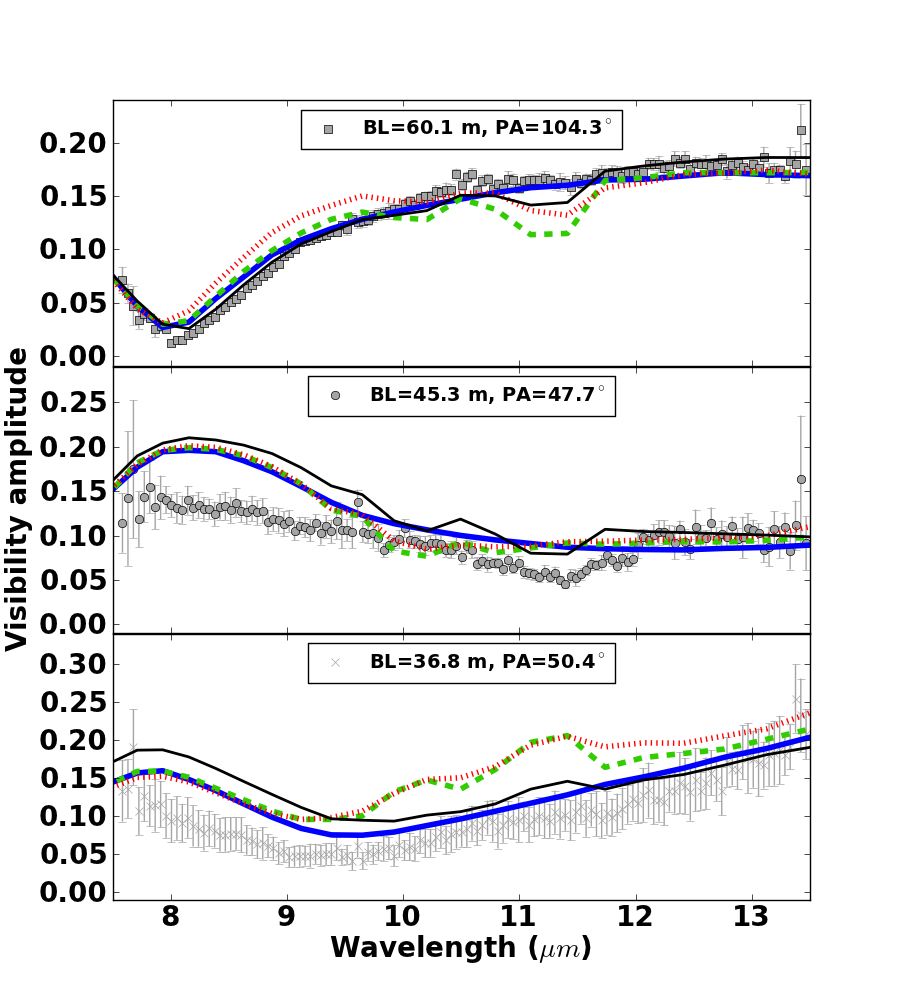}
   \caption{MIDI visibilities of AC\,Her at the three baselines (one in each panel) as a function of wavelength. 
   Overplotted are predictions for four double-power-law radiative transfer models: the standard best-fit model (full blue lines, 
   Model 2 in Table~\ref{table:MCMaxmodels}); the standard model added with 1\% of 0.1~$\mu$m 
   forsterite grains (dashed green lines); the standard model added with 1\% of forsterite grains with a size distribution 
   between 0.1 and 2.0~$\mu$m (dotted red lines); and the New Forst model from Sect.~\ref{subsubsection:Forstdensity}
   which has 1\% of 0.1~$\mu$m forsterite grains, but with a different spatial distribution (thin black lines).
     }
     \label{figure:MIDIvis-wl}
 \end{figure}

\begin{figure}
   \centering
   \includegraphics[width=\hsize]{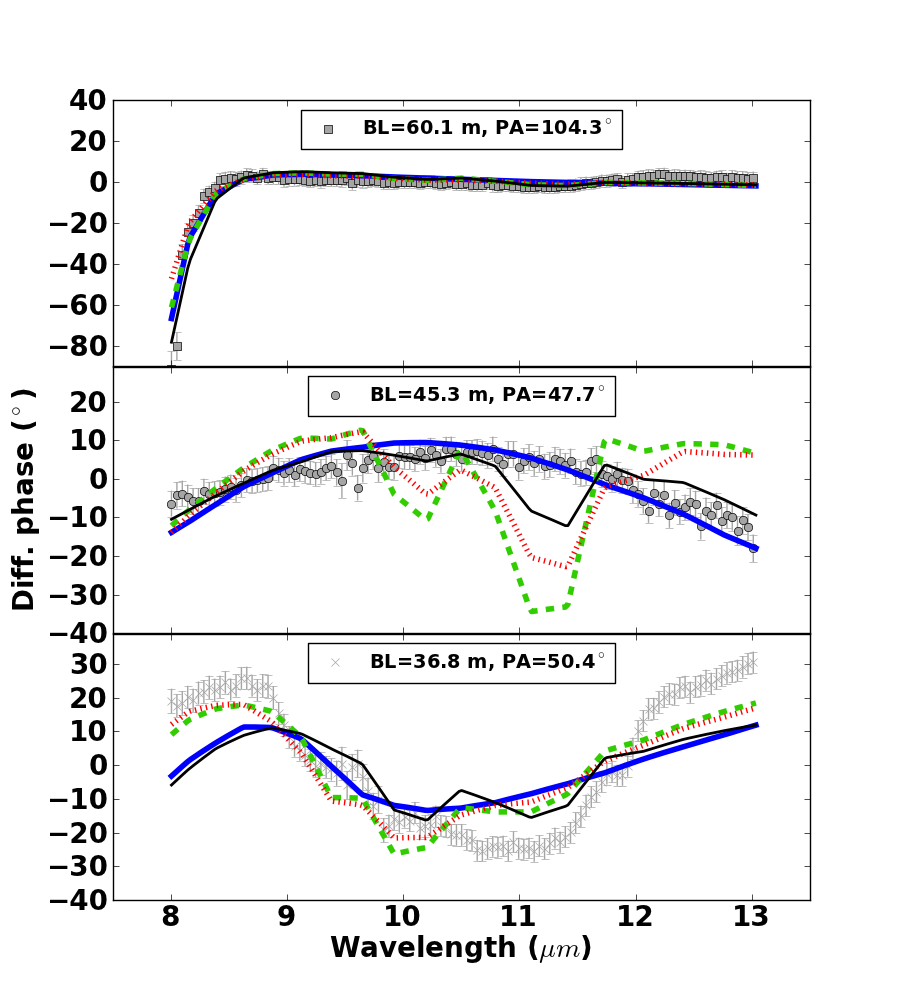}
   \caption{MIDI DPs of AC\,Her at the three baselines (one in each panel) as a function of wavelength. Overplotted are the 
   predictions for the same radiative transfer models as in Fig.~\ref{figure:MIDIvis-wl}.
     }
     \label{figure:MIDIDP-wl}
 \end{figure}

 The SED, visibility amplitudes, and differential phases following from the base model are shown in 
 Figs.~\ref{figure:fullSED},~\ref{figure:MIDIvis-wl}, and~\ref{figure:MIDIDP-wl}, respectively. A visual rendering 
 is given in the online material, in Fig.~\ref{figure:image}.
 A closer inspection of the observables shows that they all fit the observations well, except for three small 
 discrepancies. First, there is a deficit in the model SED in the 2-8~$\mu$m region. In particular the slope between
 4~$\mu$m and 8~$\mu$m is not matched by the model (see Sect.~\ref{subsubsection:fluxdeficit}). 
 Second, the predicted visibilities for the two short baselines are larger than the observed values at 8-9~$\mu$m. 
 Given the limited uv coverage one cannot, however, expect to constrain the shape of the inner rim in full detail. 
 Some of the degeneracy in inner rim shape will be disclosed in Sect.~\ref{subsubsection:extendedgrid}.
 Third, the differential phase at the shortest baseline lacks the dip near 11~$\mu$m. 
 This is to be expected because our model does not include crystalline forsterite, which is clearly present 
 in the spectrum near 11.3~$\mu$m. We examine the effect of adding forsterite to our model in Sect.~\ref{section:features}.

 \subsubsection{An alternative composition as a solution for the flux deficit?} \label{subsubsection:fluxdeficit}
 Here we describe a grid of models with an alternative amorphous 
 dust composition. We take the extreme case where there is only silicate dust consisting of amorphous
 MgFeSiO$_4$. Large particles with this composition have higher absorption efficiencies 
 in the wavelength range where we observe the flux deficit, compared to 
 particles with our standard composition (see the online Fig.~\ref{figure:OpacityComparison}).
 We want to see 1) whether our derived model parameters depend significantly on 
 our assumed dust composition, and 2) whether the 4-8~$\mu$m spectral slope mismatch can be linked to the dust 
 composition.
 
\onlfig{
 \begin{figure}
   \centering
   \includegraphics[width=8cm,height=6cm]{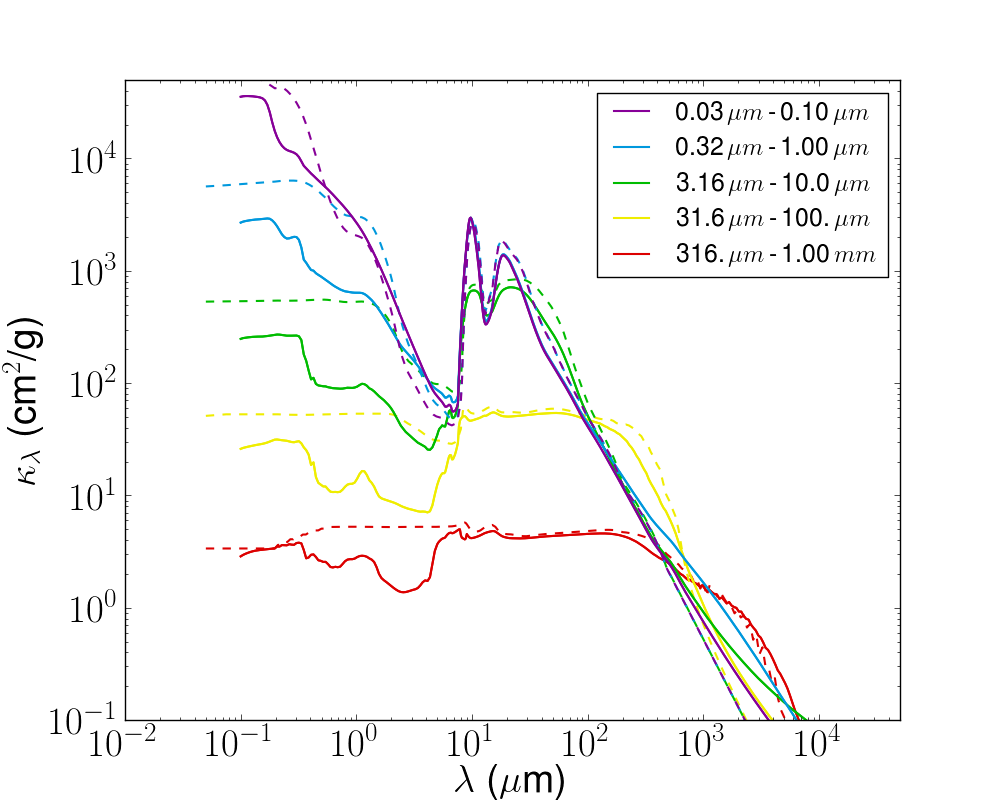}
   \caption{The mass absorption coefficients of the mixtures used in this work. Our ``standard'' mixture (the full lines) 
   contains 15\% metallic iron, while the alternative mixture (the dashed lines) consists of only MgFeSiO$_4$. 
   Only five grain sizes are shown for clarity.}
     \label{figure:OpacityComparison}
 \end{figure} 
}
 \begin{figure}
   \centering
   \includegraphics[width=8cm,height=6cm]{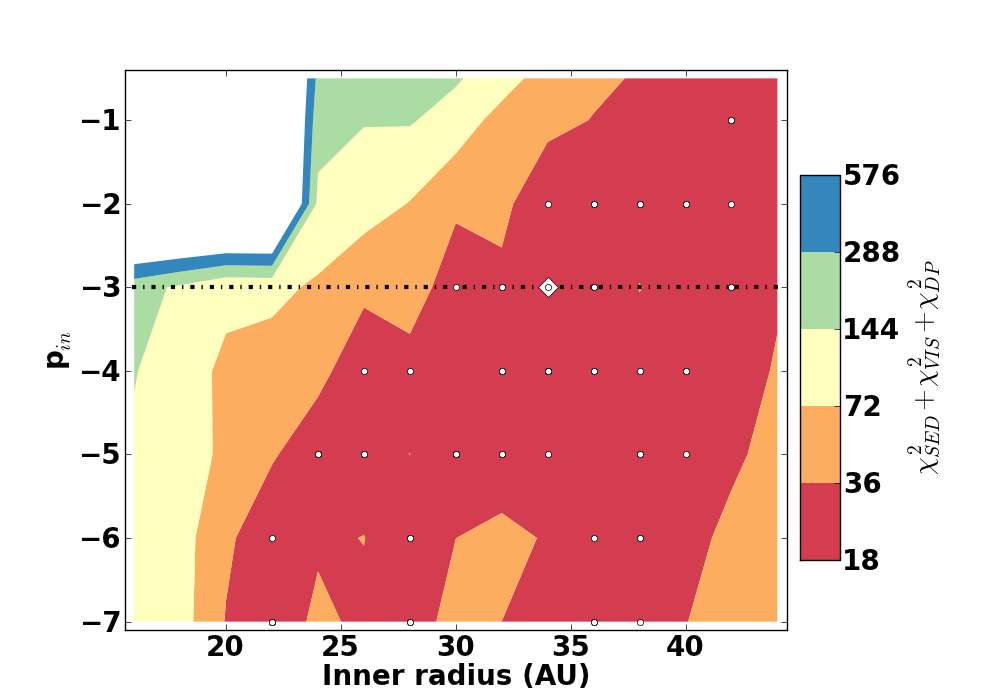}
   \caption{$\chi^2$ map of the inner radius R$_{in}$ versus the inner surface density power-law p$_{in}$. For each value 
   of these two parameters, the model with the lowest $\chi^2_{SED} + \chi^2_{VIS} + \chi^2_{DP}$ in the grid is included 
   in the $\chi^2$ map. The white dots indicate the models that were recomputed with a forsterite abundance of 1\% included.
   The white diamond denotes the location of our base model. The black dot-dashed line shows the boundary between our original 
   and the extended grid.}
   \label{figure:RinPinCHI2map}
 \end{figure}
 
 The same trends are found as for the standard models 
 previously, except that a higher inclination of 55$^\circ$ is now slightly preferred. One model 
 that stands out is incorporated as the best fit in Table~\ref{table:MCMaxmodels}. The SED of this model 
 is included in Fig.~\ref{figure:fullSED}. The flux deficit is clearly diminished, 
 but not completely removed. All observables fit better except the differential phases for which there is little change, 
 as attested by the $\chi^2$ values. 
 Although this model fits noticeably better, in the remainder of this paper we build upon our base model with the 
 standard ISM dust composition. Our main conclusion is that our derived model parameters are not very dependent 
 on the assumed dust composition.

 \subsubsection{Inner rim degeneracies}\label{subsubsection:extendedgrid}
 In Sect.~\ref{subsubsection:standardmodel} we parameterized the radial surface density distribution 
 with two power-laws in order to smooth the inner rim, which is necessary to fit the interferometric data. 
 The various parameters are, however, not all fully constrained because of the limited uv and wavelength coverage of our data. 
 The inner rim can thus appear different for similarly fitting models. Here we explore the 
 parameter space further, i.e., of p$_{\rm{in}}$, to explicitly show the degeneracies.
 
 We now extend our grid and compute models up to p$_{\rm{in}}=-7.0$. Figure~\ref{figure:RinPinCHI2map} 
 shows the resulting $\chi^2$ distribution for the extended grid, folded onto a 2D surface containing p$_{\rm{in}}$ 
 and~R$_{\rm{in}}$. So, for each value of p$_{\rm{in}}$ and R$_{\rm{in}}$, the best-fit model for all the other free parameters is included 
 in the figure. We note that for graphical purposes we take the sum of the three values of $\chi^2$. From the figure it is obvious
 that one has to be careful with the interpretation of the best-fit inner rim radius. Our data clearly exclude optically thick
 emission within $\sim$30~AU (models with p$_{\rm{in}}$ up to -3.0), but allow optically thin emission up to inner radii of 
 20~AU, when the extreme values for p$_{\rm{in}}$ are used. For such models the optical $\tau = 1$ surface in the midplane is 
 generally still located at $\sim$30-35~AU. 
  
\section{The forsterite features} \label{section:features}
In this section we describe how our base model (Best Double) looks when forsterite is included as a 
dust species. We investigate the signatures of forsterite in the mid-IR, at 69 $\mu$m and in the 
interferometric data. We do not attempt to fit the feature shapes in full detail, but merely focus on the 
relative feature strengths. 

\subsection{The base model with forsterite}\label{subsection:basicForsterite}
\begin{figure*}
   \centering
   \includegraphics[height=6cm]{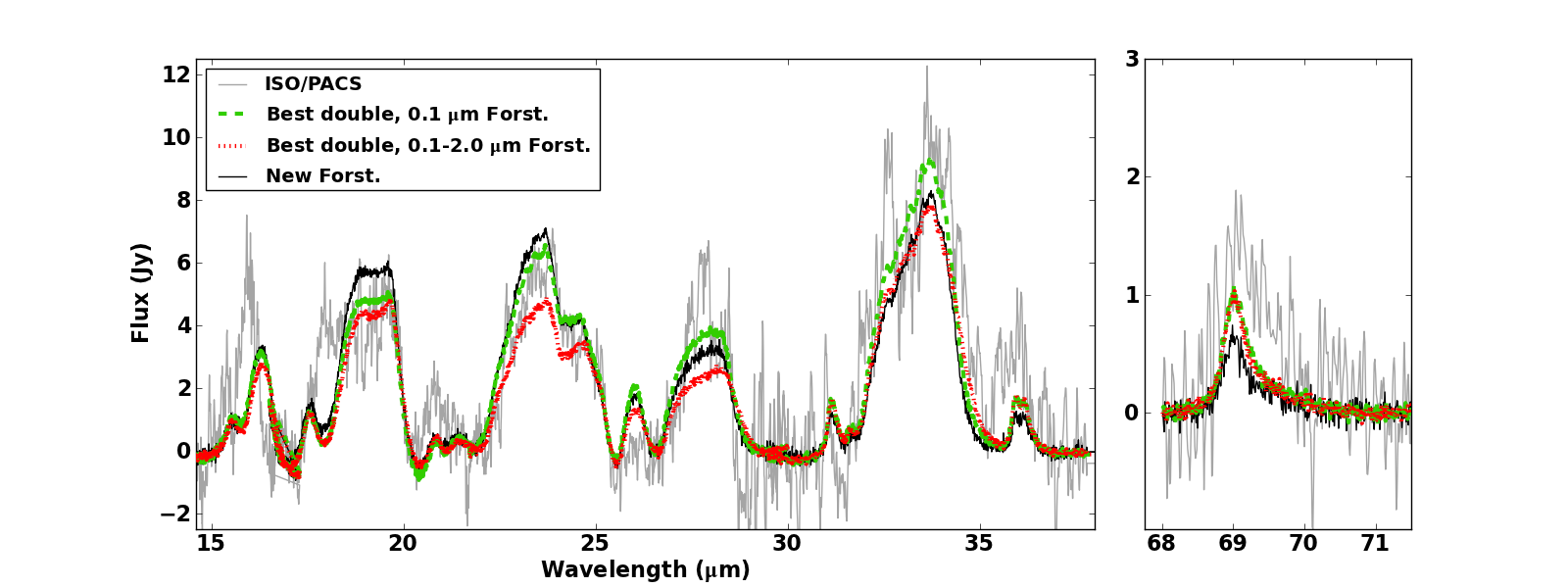}
   \caption{The forsterite features. The left panel shows the mid-IR features in the ISO spectrum; the right panel contains 
   the 69~$\mu$m band as observed by Herschel/PACS. The same models are overplotted as in Figs.~\ref{figure:MIDIvis-wl} 
   and~\ref{figure:MIDIDP-wl}, except for the base model without forsterite.}
   \label{figure:ISOforsterite}
\end{figure*}

In Fig.~\ref{figure:ISOforsterite} are shown the continuum-subtracted mid-IR spectral bands of forsterite as observed with ISO. 
In order to extract the features from the ISO spectrum we take the following wavelength regions to linearly estimate the continua (all in $\mu$m): 
14.5-15.4; 16.5-18.04; 20.1-21.9; 26.5-26.7; 29.4-31.3; 36.7-37.8; 68.0-68.4; 70.5-72.0. From blue to red wavelengths we identify 
features of forsterite at 16, 18, 24, 27, and 33~$\mu$m, and a band at 69.0~$\mu$m. The exact wavelength position of each feature is 
a function of grain size, temperature, and composition, but we use these wavelengths to refer to the different features. 

Our base model is also included in Fig.~\ref{figure:ISOforsterite}, but now added with 1\% of 0.1~$\mu$m forsterite 
grains that are distributed through the disk in the same way as the amorphous dust. 
This model reproduces the 18, 24, 27, and 33 $\mu$m bands well. 
The 16~$\mu$m band is too red in our model compared to the detection in the ISO spectrum. This was also 
reported by \citet{2007AAGielen}. Even though the peak wavelength position of the 16~$\mu$m band of forsterite is dependent on grain size, 
temperature, and composition, these parameters cannot explain its blueshifted location, as they
shift the band to redder wavelengths. 

The base model with forsterite also shows a strong 69~$\mu$m band, but it differs from the Herschel/PACS detection. The observed 
band is broader and its wavelength position redder than the one in our base model. The opacities of the 69~$\mu$m band 
significantly broaden and shift to the red with increasing temperature, which means that our model misses warm 
forsterite grains of several hundred degrees K. Such temperatures do not require it to be within the inner rim which has a temperature of $\sim$400~K. 

\begin{figure}
   \centering
   \includegraphics[width=8cm,height=8cm]{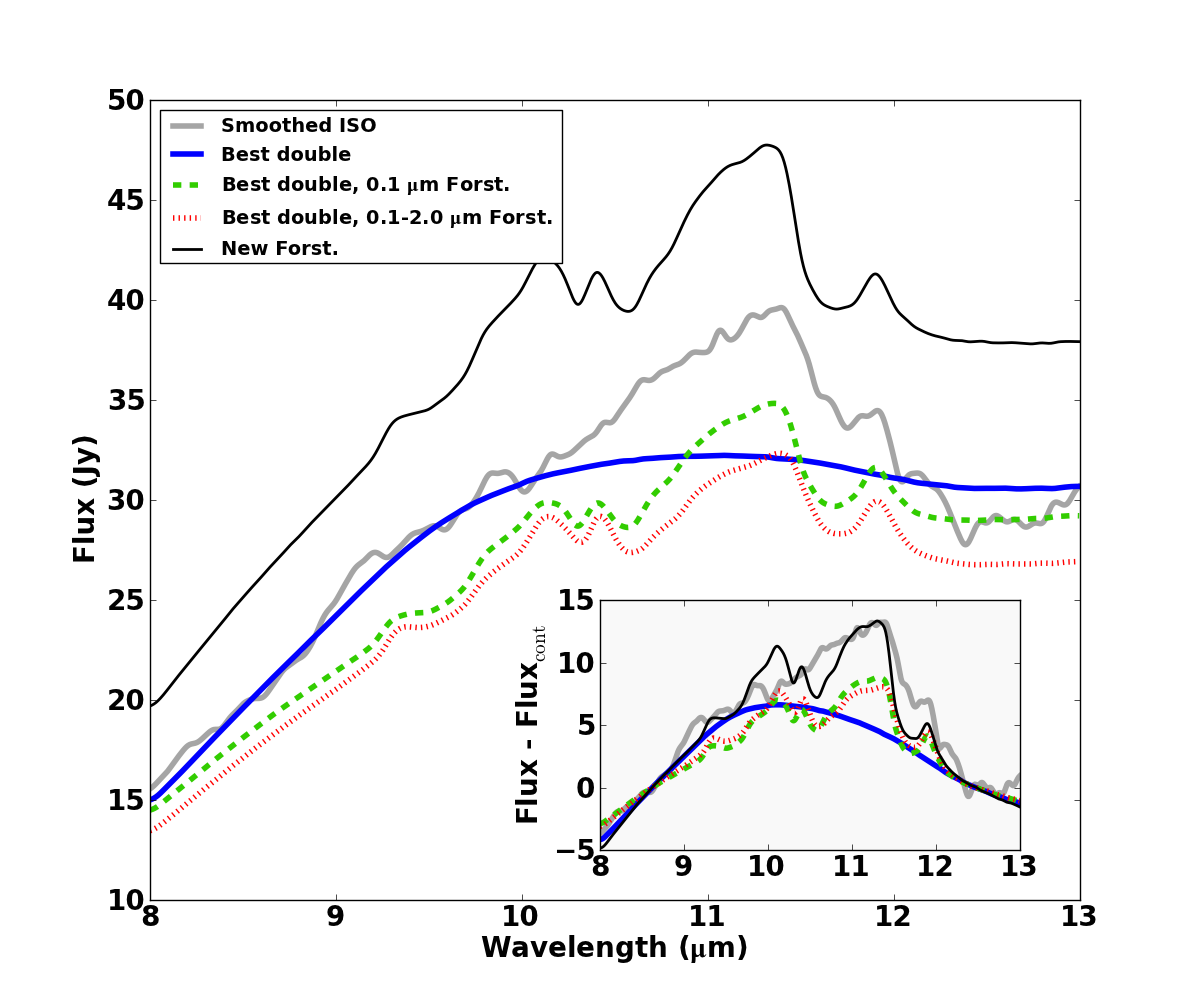}
   \caption{The 10~$\mu$m complex. The full black line is the smoothed ISO spectrum (as in Fig.~\ref{figure:MIDIDPcomparison}). 
   The overplotted models are the same as in Figs.~\ref{figure:MIDIvis-wl} and~\ref{figure:MIDIDP-wl}.
   The inset shows the continuum-subtracted spectra (using a simple linear continuum fit).}
   \label{figure:ISOsilicate}
\end{figure}

The 10~$\mu$m region also contains forsterite features; most prominently at 11.3~$\mu$m. The 10~$\mu$m region of the ISO spectrum 
is shown in Fig.~\ref{figure:ISOsilicate}. A strong and wide emission feature is seen from amorphous silicate, centered at 9.7~$\mu$m, 
with on top a clear 11.3~$\mu$m band of forsterite. The base model without forsterite reproduces the amorphous silicate band
very well. Adding forsterite to our base model introduces an 11.3~$\mu$m band on top of the 9.7~$\mu$m band (see the 
green curve in Fig.~\ref{figure:ISOsilicate}). It compares well with the measured band, but is slightly narrower. 
The addition of forsterite to our base model reduces the quality of the fit to the 9.7~$\mu$m band in Fig.~\ref{figure:ISOsilicate}. This shows 
that the other fit parameters also require minor adjustments because of the added forsterite, but such details are beyond the scope of this work.
 
As discussed in Sect.~\ref{section:simpleAnalysis}, the contribution from forsterite to the 10~$\mu$m complex can also be seen 
in the differential phases (see Fig.~\ref{figure:MIDIDPcomparison}), and weakly in the visibilities of the 45~m baseline. 
The very similar baseline-dependent behavior of the forsterite compared to the amorphous component, at first sight, suggests 
that the forsterite emission has a very similar spatial distribution to the amorphous feature emission. 

In Figs.~\ref{figure:MIDIvis-wl} and~\ref{figure:MIDIDP-wl}, the dashed green lines correspond to the visibilities and 
differential phases, respectively, of the base model to which 1\% of forsterite is added. The model distinctly shows the 
11.3 $\mu$m band in the visibilities of the 60~m and 36~m baselines, but not in the 
visibilities of the 45~m baseline, in stark contrast to the observations.
The 11.3~$\mu$m band clearly appears in the differential phase of the 45~m baseline, although the observations 
show no such feature. Additionally, the observed differential phase of the 36~m baseline contains a clear 11.3~$\mu$m forsterite 
signature, which is only very weakly present in the model.

The discrepancies can be interpreted in the following way. Because of the much higher opacity in the features compared to the surrounding 
continuum, the $\tau = 1$ surface is reached at much shallower depths into the disk. As a result, a wavelength-differential signal 
is induced in the interferometric data because the forsterite emission arises from a more compact, centrally concentrated region 
in the disk. This is illustrated by the left panel in Fig.~\ref{figure:structure}. It shows the relative distribution of 
the azimuthally integrated 11.3~$\mu$m flux as a function of the vertical and radial position in the disk. MCMax can compute these flux contributions  
as output when ray-tracing the spectrum. The continuum and feature contributions are shown separately in the upper 
and lower panels, respectively. The continuum contribution is computed by linear interpolation from the flux distributions at two 
neighboring continuum wavelengths. The feature contribution is then obtained by subtracting the continuum contribution from the total.
Even though the low-intensity tail in the feature is somewhat larger than in the continuum (e.g., the dark blue contour), 
the region where most of the flux originates is more compact and at higher disk layers in the feature than in the continuum. The difference in flux 
distribution seems to be rather subtle (considering the scales), although the effect pops up distinctly in the observables.

\paragraph{Summary}
We can summarize the discussion of the forsterite signatures in this section by saying 1) that our model shows strong features in the 
visibilities while the observations do not; 2) that the 11.3 $\mu$m band is strongly present in the model differential phase 
of the 45~m baseline and only weakly in that of the 36~m baseline, which is opposite to the observations; and 3) that the 
spectrum is well fit over a wide wavelength range, except for the red component in the 69~$\mu$m band. Our model thus needs 
corrections. 
The 69 $\mu$m band requires more warm forsterite, which could be added to the model by increasing the forsterite abundance in 
the inner part of the disk. The interferometry on the other hand indicates that the hot forsterite in our model gives an emission 
distribution that is too centered; the 11.3~$\mu$m emission needs to be more extended than is currently the case in our base model. 
Although it is outside the scope of this paper to perfectly reproduce 
all forsterite features, we will show in the next section the effect of several parameters.

\subsection{Effect of model parameters on the features}\label{subsection:forstparstudy}
Adding forsterite to our base model does not have the desired effect on the 
interferometric observables. We identified the discrepancy as the result of an optical depth effect. Here 
we briefly explore two ways to change the optical depth in the innermost disk regions at 11.3~$\mu$m.
First we will decrease the forsterite opacity at 11.3~$\mu$m by increasing the size 
of the crystals. Second, we will explore whether the degeneracies in inner disk structure 
can help to solve the forsterite problem. 

\subsubsection{The forsterite grain size}
Grain size has the effect of weakening, shifting, and broadening the spectral features of crystalline dust grains 
\citep[see, e.g.,][]{2015AAMaaskant}. The forsterite added to our base model (Sect.~\ref{subsection:basicForsterite}) 
had a grain size of 0.1~$\mu$m. If instead we use forsterite with a grain size distribution 
from 0.1 to 2.0~$\mu$m (with a power-law slope of -3.5), then the opacity in the 11.3~$\mu$m band drops by a factor of two. 
This opacity reduction decreases the optical depth at 11.3~$\mu$m in the inner disk. 
As can be seen in Figs.~\ref{figure:MIDIvis-wl} and~\ref{figure:MIDIDP-wl}, this decline in optical depth diminishes the 
forsterite signal in the visibilities and differential phases (the red dotted line), but not sufficiently. 
The differential phase signal at the 36~m baseline remains unaffected, and the 
faulty signal at the 45~m baseline is only slightly diminished. The mid-IR spectral features (Fig.~\ref{figure:ISOforsterite}) are 
not changed much, although the 24, 27, and 33~$\mu$m complexes become slightly weaker. 
The 69 $\mu$m band remains unchanged, as expected.

\subsubsection{The inner disk density distribution} \label{subsubsection:Forstdensity}
Another way to improve the forsterite signatures may be to alter the inner density structure of the disk. 
The forsterite emission at 11~$\mu$m arises solely from the innermost disk regions and may be affected 
by our particular choice of base model. The optical depth profiles may be quite different at 11~$\mu$m between the 
various models included in Fig.~\ref{figure:RinPinCHI2map}. After all, one must not forget that the parameters 
of our base model were determined from a fit to the interferometric data in which forsterite was
neglected.

For all the models within the extended grid computed in Sect.~\ref{subsubsection:extendedgrid} for which each 
of the values of $\chi^2$ is smaller than $5 \times$~$\max(1.0,\chi^2_{\mathrm{min}})$, 
we now recompute the visibilities and differential phases with 1\% of 0.1~$\mu$m grains of forsterite included. These models 
are indicated as white dots in Fig.~\ref{figure:RinPinCHI2map}.
From this newly computed set, models with very optically thin inner rims (p$_{\rm{in}} < -4.0$ and R$_{\rm{in}} < 30$~AU) do not lead to a
significantly improved interferometric forsterite signature, nor do the more extended rims (R$_{\rm{in}} > 36$~AU). 
For the following qualitative discussion we select the model that matches the differential phases best, 
since the immediate effect of the forsterite on the fit quality is more apparent in the differential phases than in the visibilities.
This model is included in Figs.~\ref{figure:MIDIvis-wl},~\ref{figure:MIDIDP-wl},~\ref{figure:ISOforsterite},
and~\ref{figure:ISOsilicate} as the thin black full line (referred to as ``New Forst'' in the legends). 
For this model, a small signal appears in the 36~m baseline measurement,
while the erroneous signal at the 45~m baseline is significantly diminished. The visibility signals are also 
reduced to an acceptable level. At the same time the mid-infrared features are little affected. Only the 69 $\mu$m fit is 
worse: the blue part of the band becomes less strong, and the broad red-shifted component is still not reproduced.
Finally, we note that the amorphous contribution to the 10~$\mu$m complex is too strong for this model. 

\begin{figure*}
   \centering
   \includegraphics[width=8cm,height=10cm]{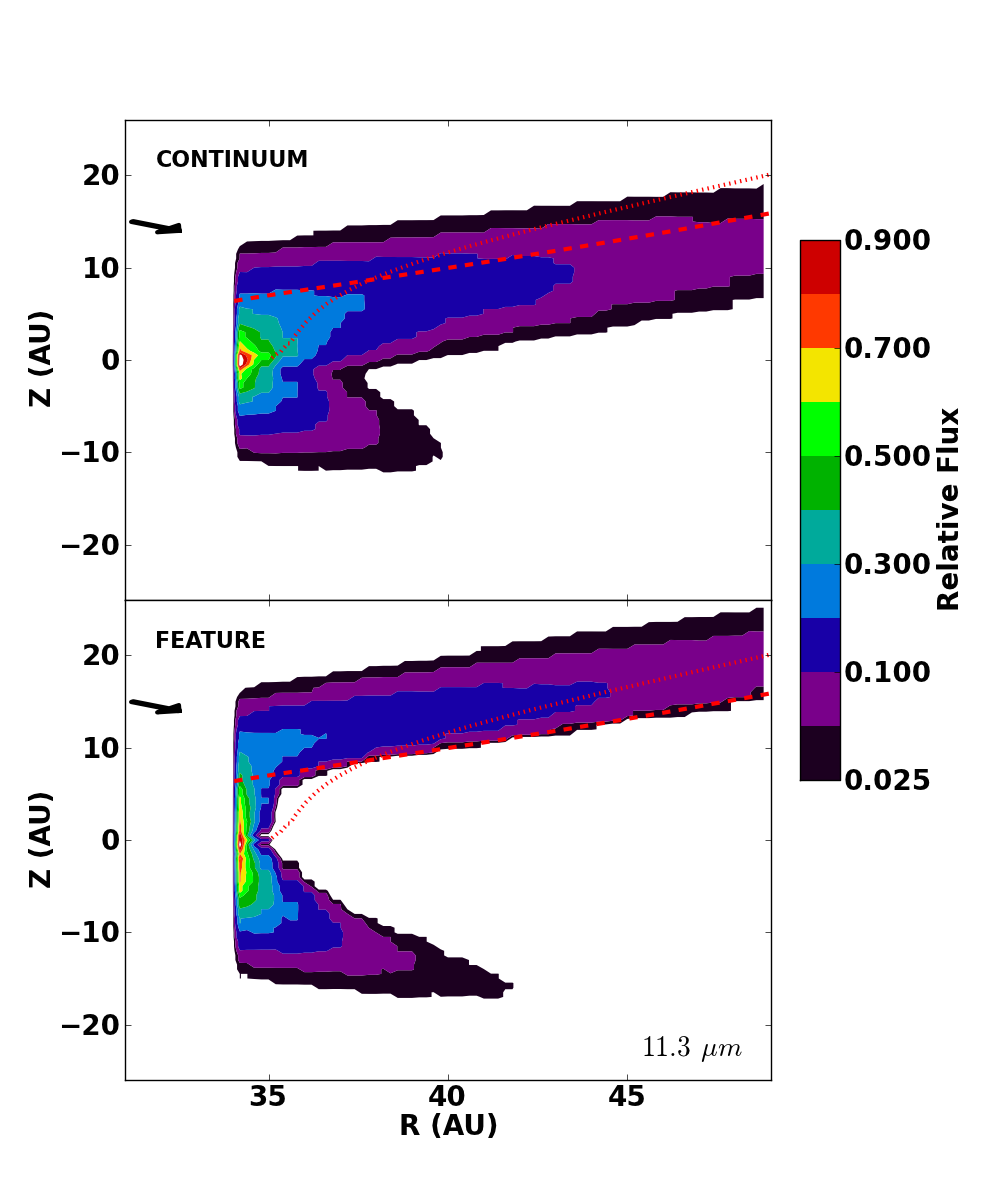}
   \includegraphics[width=8cm,height=10cm]{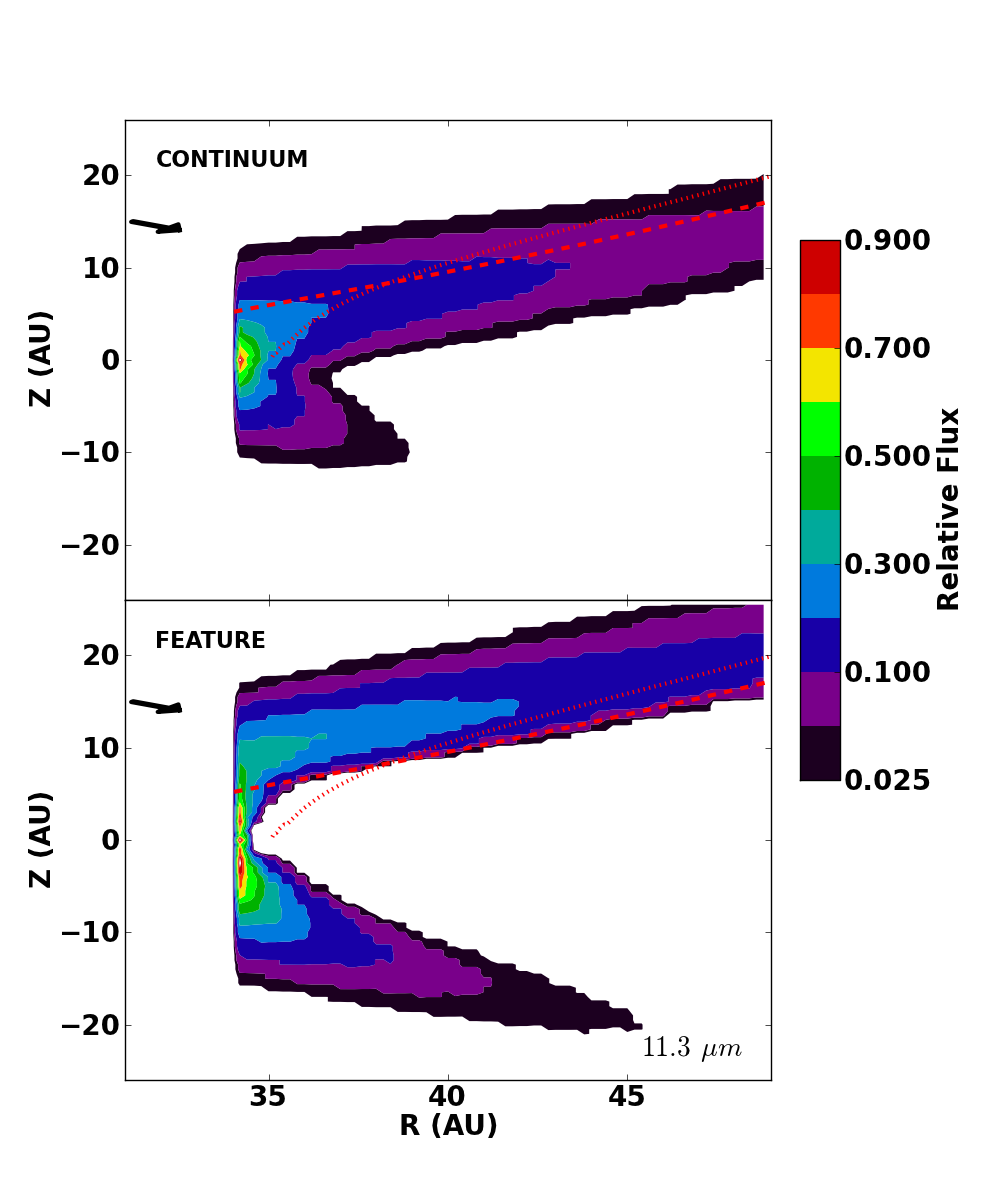}
   \caption{Flux contribution at 11.3~$\mu$m as a function of radius and height in the disk, for the continuum (from interpolating 
   nearby continuum wavelengths) and the feature. The red dashed and dotted lines represent the vertical and radial $\tau = 1$ surfaces
   at 11.3$\mu$m, respectively. Left panel: the base model with 1\% of 0.1~$\mu$m forsterite grains. 
   Right panel: the ``New Forst.'' model explained in Sect.~\ref{subsection:forstparstudy}. The arrow indicates the viewing direction.}
   \label{figure:structure}
\end{figure*} 

We show the flux distribution at 11.3~$\mu$m in the right panel of Fig.~\ref{figure:structure}. 
This model fits the interferometric data better than the base model 
with forsterite because the forsterite emission is now distributed over a more 
extended area (see in particular the light blue region in the upper part of the disk). 

This is the result of the combined effect of the three parameters that have changed compared to the base model:
\begin{itemize}
 \item the decrease in maximum grain size from 1.0 to 0.3~mm causes an increase in the dust mass of small particles (for the same 
 total dust mass and power-law index of the grain size distribution), hence an increase in the continuum optical depth.
 The forsterite remains unaffected by this parameter.
 \item the steeper p$_{\rm{in}} = -4.0$ results in a decreased density, and hence optical depth, at the inner rim. This parameter
 also shifts some forsterite from the inner regions in the direction of the turnover radius. Hence, the abundance in the upper disk layers 
 increases at radii $\sim$40~AU, and thereby also the emission from these regions.
 \item the decreased inclination of 45$^\circ$ causes the viewing direction to be more vertical. As a result, the integrated optical 
 depth in the line of sight towards farther-out disk radii decreases, so that more photons can 
 escape these regions in the direction of the observer (compare the regions delineated by the vertical and 
 radial $\tau=1$ surfaces included as red dotted and red dashed lines in Fig.~\ref{figure:structure}, respectively).
\end{itemize}


In summary, although the New Forst model fits the interferometric data better than our base model with 1\% of forsterite, 
this model does not reflect the global structure of the disk better than our base model (e.g., the mm fluxes are now too small). 
The new parameter values do, however, alter the spatial distribution of the forsterite emission in the innermost regions, differentially 
with respect to the continuum.
The optical depth due to the forsterite is decreased in the innermost upper disk regions, without too much change to the continuum
at these wavelengths. This model illustrates how the spatial distribution of the forsterite emission should change to fit the 
interferometric data better.
We conclude that the details of the inner rim density structure may subtly affect the distribution of the forsterite emission, 
and thereby the features in the interferometric data.
  
\section{Discussion} \label{section:discussion}
In this paper we have clearly demonstrated that the mid-IR brightness distribution of the AC\,Her system appears asymmetric, and that this 
is due to the circumstellar material being in a passive dusty puffed-up circumbinary disk that we see under 
a well-constrained inclination of 50$\pm$5$^\circ$. We have shown this by applying state-of-the-art Monte Carlo radiative transfer disk models in 
hydrostatic equilibrium to three interferometric measurements from the MIDI instrument on the VLTI of very high quality, combined with 
the full SED and the mid- to far-IR spectral features of crystalline forsterite. Overall we find a very good 
agreement between our best-fit models and the observables. To accurately fit the MIDI visibilities, and in particular 
the differential phases at the different baseline position angles, requires a highly asymmetric 
apparent source morphology which is, as we have shown, naturally generated by a disk model. 

The inclusion of mm-sized grains 
that are allowed to settle to the midplane of the disk leads to a very good match with the full measured SED, just like for the 89 Her 
system \citep{2014AAHillen}. Although there is still a flux-deficit in the 2-8~$\mu$m wavelength range, 
we argued that the discrepancy is within the uncertainties of our models and assumptions.
We investigated this by putting into question the assumed composition of the amorphous dust component,
but there are other possible origins. For one, our models ignore the effect of molecular opacity on the thermal structure 
of the disk (as well as the resulting direct emission contribution to the spectrum). \citet{2014ApJMalek} claim this 
to be significant for the similar HR4049 system.
Similarly, optically thin dust might be present within the inner hole, e.g., in an accretion 
flow. Our exploration of inner rim shapes in Sect.~\ref{subsection:forstparstudy}, 
and the degeneracies with respect to the current data, shows that there are still several 
degrees of freedom. 

Our results, nevertheless, resolutely confirm the disk hypothesis for the AC\,Her system that was raised by 
\citet{1998AAVanWinckel} based on less direct arguments. AC\,Her is, moreover, one of the objects among the sample of suspected
post-AGB binaries with disks of \citet{2006AAdeRuyter} for which the near-IR excess is least pronounced. 
Therefore, our results corroborate that most of the 
objects listed by \citet{2006AAdeRuyter} must indeed be disks as well.

Our results, however, also raise several questions. 

\subsection{The inner hole}
One of the most striking results of this paper is the large inner radius of the circumbinary disk in AC\,Her.
The typical $\tau = 1$ radius in the midplane of the disk is in the 30-35~AU range, irrespective of any speculative inner optically 
thin contribution. Following Eq.~12 of \citet{2010ARAADullemond}, one can simply estimate 
the dust sublimation radius expected for the stellar parameters of AC\,Her, which we find to be in the range of 1.5-5.0~AU, 
depending on the assumed value of their $\epsilon$. This parameter incorporates all inner rim properties that 
have an influence on the evaporation of the dust (like grain size and composition). We thus find that the actual inner rim in the 
AC\,Her system lies about an order of magnitude beyond the expected dust condensation radius. We note that both radii 
are dependent on the adopted distance, so their ratio is independent of it.

AC\,Her is the first post-AGB disk system for which it is now firmly established that the inner disk radius 
is significantly beyond the dust condensation radius. One can only speculate about the physical origin 
of this large inner hole. Several mechanisms have been proposed in the literature for the formation of holes or gaps
in protoplanetary disks, such as photoevaporation, accretion, or the dynamical influence of a planetary-mass companion.
More systems like AC\,Her need to be studied with high angular resolution techniques and across the full electromagnetic spectrum 
to show which of these mechanisms are at work in post-AGB disks.

\subsection{The total disk mass and its composition}\label{subsection:abundances}
One of the parameters that is fairly well constrained in our modeling is the dust mass thanks to the inclusion of the 
mm fluxes. In each of the grids our best-fit dust mass is larger than 10$^{-3}$~M$_\odot$. In fact, the only model 
in which the dust mass is smaller than $2.5 \cdot 10^{-3}$~M$_\odot$ is the best-fit single-power-law model, which 
clearly lacks emission in the mm regime. In summary, under the assumptions made in our modeling (in particular 
the composition of our dust, p$_{\mathrm{out}} = 1.0$, R$_{\mathrm{out}} = 200$~AU, and $\alpha = 0.01$), 
the minimum dust mass needed to fit the SED is $\geq$10$^{-3}$~M$_\odot$. 

If the gas/dust ratio were 100, then the total disk gas mass would be 0.1~M$_\odot$. Our modeling indicates, 
however, that the global gas/dust ratio is likely smaller. Although our best-fit model has a ratio of 10.0, 
there are models that formally fit our acceptance criteria with a ratio of 1.0.  
We come to the conclusion that the total gas mass must be at least 10$^{-3}$~M$_\odot$, and can be one or two orders of 
magnitude larger, depending on the theoretically ill-understood parameter $\alpha$. 
We note that $\alpha$ and the gas/dust ratio have a similar influence in our models \citep[see, e.g.,][]{2012AAMulders}.
Interestingly, the total gas mass found for AC\,Her by \citet{2013AABujarrabalB} is 
$\sim$8$\cdot10^{-4}$~M$_\odot$. If we correct for their smaller adopted distance, their gas mass is about a factor of five 
smaller than our best-fit value. This is within the respective errors of both methods. 
It would nevertheless be useful to examine how much their results may be 
influenced by optical depth effects. In addition, resolved observations with ALMA would allow us
to constrain some of the model parameters assumed in this work, like p$_{\mathrm{out}}$ 
and R$_{\mathrm{out}}$, even though we do not expect them to significantly alter our results. 
 

One assumption that may significantly affect our derived mass, 
is the dust composition. Although we have checked and ruled out a serious influence 
of the chosen silicate (+metallic Fe) composition, one cannot exclude other dust species with a high (sub-)mm opacity 
that would lower the effective dust mass. For protoplanetary disks a component 
of amorphous C is often assumed. There are claims that in certain post-AGB disks 
amorphous C may be present as well \citep[e.g.,][]{2013AAAcke}. \citet{2014AAHillen}
showed that for the 89 Her disk, models with amorphous C can fit the observations equally well. The 
presence of C-rich dust in an otherwise O-rich environment, however, requires an explanation.

Finally, one can do an interesting exercise to check how realistic our derived dust 
mass is, keeping in mind that all the circumstellar material has its origin in the central post-AGB star. Assuming that 
the post-AGB star started its evolution with a solar metallicity, one can compute how much mass 
of Fe, Si, and Mg the star has lost throughout its evolution. Assuming the star lost 1~M$_\odot$ of 
hydrogen, we arrive at $\sim$1.8, 0.9, and 1.0$\cdot 10^{-3}$~M$_\odot$ of Fe, Si, and Mg, respectively.
Assuming that the disk's current dust mass is indeed 2.5$\cdot 10^{-3}$~M$_\odot$, and taking our standard 
dust composition, then about 0.5$\cdot10^{-3}$~M$_\odot$ of Fe, Si, and Mg would be currently locked up 
in dust in the disk. This means that between 30 and 50\% of all the Fe, Si, and Mg that was released by the 
star should still be present in the form of dust in the disk. Such large fractions seem unrealistically high, 
as they would imply that the disk mass had once been close to 0.5~M$_\odot$, and definitely warrants 
further investigation. All potential sources of bias mentioned previously should be checked 
thoroughly. 

\subsection{Forsterite}
AC\,Her joins the short list of disk objects for which the mineralogy of crystalline dust species like forsterite 
has been studied with interferometry. Although rare, such studies are powerful.
Based on a few MIDI observations, \citet{2004NaturVanBoekel} for the first time directly measured a
gradient in the chemical composition of the dust in several protoplanetary disks. 
Using the interferometric correlated fluxes as a proxy for the inner disk spectrum, they  
fitted the mineralogy in the inner and outer disk separately. \citet{2006AADeroo} later did a similar 
analysis for the post-AGB system HD52961, and found that the crystalline dust in that disk is also concentrated 
in the hotter inner regions. These disks were less resolved than in the current work, so 
their analysis was not complicated by the detailed morphologies of the objects. The few interferometric 
measurements analyzed in this paper are, however, dominated by the geometry, so that radiative transfer models are 
needed to draw conclusions concerning the morphology and mineralogy of the disk. 
We have shown in this paper that wavelength-differential interferometric observables have a strong diagnostic value in order
to constrain the spatial distribution of minerals thanks to their high reliability and resolving power. Even without the presence 
of a signal (see, e.g., our longest baseline), constraints can be derived.


We find that by including forsterite with a similar radial distribution throughout AC\,Her's disk to the amorphous dust and
a global abundance of 1\%, the mid-IR features of crystalline forsterite are well reproduced. 
This model does not reproduce the 69 $\mu$m band well, nor the forsterite signatures in the MIDI 
data. The observed 69 $\mu$m band shows that there must be an enhanced abundance of warm 
forsterite of several hundreds of kelvin. Based on our parameter study in Sect.~\ref{subsection:forstparstudy}, 
we conclude that the interferometric data require a similar change to the forsterite emission at 
11.3~$\mu$m: less emission from the hot innermost AU of the disk and more emission from the warm forsterite in the 
upper disk layers a few AU farther out (See Fig.~\ref{figure:structure}).

\subsection{Comparison with protoplanetary disks}
In this section we compare AC\,Her with some recent results concerning protoplanetary disks. The most extensive statistical and 
purely empirical comparison between the galactic post-AGB disks and a set of protoplanetary disks was done by \citet{2006AAdeRuyter}.
They used the empirical separation in two groups of Herbig AeBe disks determined by \citet{2004AAAcke} to show that all post-AGB disks 
are similar to those of the second group. These group II sources are generally believed to be flatter 
and more continuous disks than the group I objects, which have much more pronounced and redder far-IR excesses due to their 
flaring upper layers. Recently, \citet{2013AAMaaskant} concluded that all group I 
sources are indeed likely flaring disks with gaps. It is interesting to note that AC\,Her is one of the post-AGB objects that 
is most closely located to the dividing line between the group I and II sources. The inner hole in AC\,Her is  
in absolute scale similar in size to the largest gaps found in the small sample of 
group I Herbig sources by \citet{2013AAMaaskant}. 
However, in terms of the R$_{\mathrm{in}}$/R$_{\mathrm{cond}}$ ratio, AC\,Her's hole is
smaller than the gaps in the Herbig sources as a consequence of the large difference in stellar luminosity. 
A very relevant question is, why are there no group I post-AGB disks? In addition to the presence of a gap, 
the defining characteristic of a Herbig group I source seems to be a high degree of flaring. Our results combined with the 
statistics from \citet{2006AAdeRuyter} suggest that, although there are post-AGB disks with large inner holes, the presence 
of holes does not seem to go hand in hand with a large degree of flaring.

Next we compare our results concerning the forsterite in AC\,Her to the protoplanetary disk sample studied by \citet{2015AAMaaskant}.
Using the observations from the Herschel DIGIT programme, \citet{2015AAMaaskant} analyzed a sample of Herbig disks (containing both
group I and II sources) to search for trends in the forsterite emission characteristics and study how these compare with typical 
radiative transfer models assuming various disk evolution scenarios. One of their main results concerns the relation between the 
69~$\mu$m band luminosity and the integrated feature strength ratio of the 23 over the 69~$\mu$m bands (their Fig.~3). They find 
a clear separation in forsterite behavior between group I and II sources. Group I objects are scattered in this diagram, 
but typically show larger 69~$\mu$m band strengths. Group II objects on the other hand have very weak 69~$\mu$m bands, often
below the detection threshold of the PACS instrument, even though the corresponding 23~$\mu$m features are well detected in the Spitzer spectra.
\citet{2015AAMaaskant} conclude that the weak 69~$\mu$m bands in group II sources are the result of inefficient radial mixing
in such disks, so that they only contain warm forsterite.

The integrated feature strengths of AC\,Her (L$_{69}$/L$_\star \sim 2.6 \cdot 10^{-5}$ and I$_{23}$/I$_{69}\sim75$) 
make it falls in a region only populated by group I protoplanetary disks. Although some group II Herbig sources have similar 
I$_{23}$/I$_{69}$ ratios, high 69~$\mu$m luminosities only seem to be reached by group I objects. Interestingly, 
out of the six post-AGB disk sources targeted with Herschel/PACS, the 69~$\mu$m band was only detected in 
AC\,Her and the Red Rectangle \citep{2014AABlommaert}, which both have peculiar disks. The sample of \citet{2014AABlommaert} 
combined with our detailed modeling of AC\,Her 
confirms the link between the presence of a detectable 69~$\mu$m band and the inner structure of the disk, as found 
by \citet{2015AAMaaskant} and following the work of \citet{2011AAMulders}.



\citet{2011AAMulders} fitted for the first time the mid-IR features and 69~$\mu$m band of a disk source, HD100546, with detailed 
radiative transfer models to investigate the influence of optical depth effects arising from the disk structure.
They showed that the mid-IR bands can be reproduced with a forsterite abundance of 30\% at the inner wall
and falling as r$^{-2}$ throughout the inner half of the disk. For this model the 69~$\mu$m band is, however, too narrow and its wavelength 
position too blue compared to the observations. They varied the spatial distribution of the forsterite in various ways
and found that a model with a very high (40-60\% by mass) local abundance of forsterite in the inner wall and 
inner disk surface layers can reproduce the mid-IR features and the 69 $\mu$m band simultaneously.
Our forsterite modeling highlights the similarity with the forsterite features of AC\,Her, although the L$_{69}$ is lower.  
The forsterite abundance in the inner disk surface layers of AC\,Her might thus be similarly enhanced as in HD100546.
It is beyond the scope of this paper to investigate such a model in detail for AC\,Her. Interferometric observations 
with a better uv coverage would be needed to constrain the details.
It would be worthwhile to do a similar investigation of the 
forsterite signatures in the interferometric data of HD100546. An extensive MIDI data set exists 
for this object \citep{2013AAMuldersB,2014AAPanic}. Our results confirm the suggestion by \citet{2011AAMulders}
that the strength of crystalline features may be a reflection of the spatial distribution of the forsterite in the disk, 
intimately related to the structure of the disk, rather than the overall crystal abundance.


A higher abundance of forsterite in the inner disk, just like in several protoplanetary disks, would 
imply that the forsterite in AC\,Her's disk is largely created as the result of 
disk processing, after the disk formed. Nonetheless, the current inner radius of AC\,Her's disk at $\sim$34 AU makes it 
unlikely that the observed crystalline material was assembled where it is presently observed.
A larger and more quantitative exploration of possible spatial distributions of the forsterite needs to be 
performed before a final conclusion can be drawn about its formation history.
Our finding of an enhanced abundance in the inner disk nevertheless suggests that it may not have formed 
alongside with the amorphous dust when the disk was created.

\subsection{Evolutionary constraints} \label{subsection:evolution}
Finally, using our newly derived constraints one can ponder the evolution of the AC\,Her system into its current state. 
Adopting the PARSEC stellar evolutionary tracks \citep{2012MNRASBressan}, one finds that our derived 
stellar luminosity is reached at the tip of the red giant branch for a star with initial mass M$_i \sim$1.0-1.4~M$_\odot$.
This makes AC\,Her a candidate to be a Galactic counterpart to the rather abundant post-RGB disk systems found in the 
Small Magellanic Cloud by \citet{2014MNRASKamath}. This conclusion is only tentative, however, given the large luminosity error 
due to the significant spread in the PL-relation.

Assuming then the $\log g$ and stellar radius as listed in Table~\ref{table:parameters}, one finds a primary mass of 
$0.6 \pm 0.2$~M$_\odot$. Combining this value with our interferometrically determined 
inclination (assuming co-planarity between the disk and the binary) and the orbital solution of \citet{1998AAVanWinckel}, we find a 
companion mass of $1.2 \pm 0.2$~M$_\odot$, or a M$_1$/M$_2$ mass ratio of $0.5 \pm 0.1$. The primary's Roche lobe radius 
is then 180$\pm$20~R$_\odot$, which means that the system is currently not interacting. This is consistent with the lack of 
observational evidence of ongoing activity. 

The radius that the primary would have had at the tip of the RGB ranges 
between 170 and 150~R$_\odot$, which is remarkably similar to the current size of the Roche lobe. This suggests that the 
disk in AC\,Her was indeed formed by Roche-lobe overflow. Where all the envelope mass of the primary went is not clear.
The secondary cannot have accreted much more than $\sim$0.2~M$_\odot$ because otherwise the initial mass ratio would 
have been so high that the Roche-lobe overflow would have been unstable and the system would have evolved into 
a common envelope scenario. 

In a series of analytical calculations and smooth particle hydrodynamics 
simulations, \citet{1991ApJArtymowicz}, \citet{1994ApJArtymowicz}, and \citet{1996ApJArtymowicz} have shown various 
effects resulting from interactions between circumbinary disks and their central binaries. These studies found 
that the resonant interactions can lead to the formation of a stable inner hole in the disk with a size up to $\sim$3 times the
binary separation. For AC\,Her this would be at maximum $\sim$8~AU, which is beyond the dust sublimation 
radius but well within our inner rim radius. \citet{1996ApJArtymowicz} later found that for thicker disks
than in the original simulations, accretion streams can exist within the otherwise stable hole. 
Can such accretion streams be responsible for the large inner hole in AC\,Her's disk, possibly induced by the large-scale 
luminosity variations throughout the pulsation cycle? 

\section{Conclusions}
In this paper we have extensively analyzed the circumstellar environment of the AC\,Her post-RGB/AGB binary system. 
First, we carefully collected pulsation-phase dependent photometric data to get a precise determination of the 
stellar parameters at the pulsation phase where our selected near- and mid-IR data were obtained. Given the large pulsation
amplitudes of RV Tauri variables this is essential in order to reliably model the circumstellar environment.
A basic analysis of our mid-IR interferometric data showed that the circumstellar material 
needs to be present in a globally asymmetric form that is moreover compact. This is consistent with the 
current state of the art concerning these objects which affirms that they have gas and dust in Keplerian circumbinary disks.
With a detailed radiative transfer analysis we firmly established this conclusion: a (semi-)stable circumbinary disk in 
hydrostatic equilibrium can explain all the existing observational constraints on this system. In this work we focus on 
a simultaneous fitting of the full SED and our mid-IR interferometric data, which succeeds very well by including a rounding off 
of the inner disk rim. 

We determined disk parameters like dust mass, inner radius, maximum grain size, inclination and position
angle on the sky, and found our results to be robust against several of our modeling assumptions. 
We found that the disk in AC\,Her is seen under a significant inclination, that it has a rather high total dust mass
and a large dust-poor inner hole. These derived values raise several questions concerning the current state and 
the evolution of the system. Although the precision on our distance and luminosity estimate is insufficient 
to make very strong statements, the newly derived stellar and disk parameters at least suggest that 1) AC\,Her 
may either be a post-RGB or a post-AGB star, 2) the circumbinary disk is in an evolved state as shown by the 
large inner radius and low gas/dust ratio, 3) the disk in AC\,Her used to be very massive, and 4) 
the system may be an exemplary case of binary-disk (resonant) interaction.

In addition to the global structure of the disk in AC\,Her, we also considered its mineralogy. The MIDI spectral range
includes the 11.3~$\mu$m feature of crystalline forsterite, which also makes these data a powerful probe of the spatial 
distribution of the crystalline material in the disk. In combination with the mid-IR and the 69~$\mu$m spectral features,
we have found evidence in favor of an enhancement of the forsterite abundance in the inner disk surface layers 
compared with a distribution that is similar to the amorphous dust. Based on the features alone, similar modifications have been proposed for 
the forsterite distribution in protoplanetary disks, most notably for the well-studied star HD100546. Although our results
are inconclusive, we have clearly demonstrated the value that 
interferometric data can have for mineralogical studies. With a more extensive filling of the uv-plane, such as the 
MATISSE 2$^{\mathrm{nd}}$ generation VLTI instrument will be able to provide, the spatial distribution of crystalline material
in disks can be studied more efficiently and in much greater detail. Ultimately this will give an interesting and unique 
perspective on the formation and evolutionary history of post-AGB disks.


\begin{acknowledgements}
MH and HVW acknowledge support from the Research Council of the KU Leuven under grant number GOA/2013/012.
We acknowledge with thanks the variable star observations from the AAVSO International Database contributed by observers 
worldwide and used in this research. J.M. is funded as a PhD fellow of the Research Foundation Flanders (FWO). We thank Rens 
Waters for the discussions about this research. M.H. also thanks Sara Regibo for her help with the SPIRE data.
\end{acknowledgements}

\bibliographystyle{aa}
\bibliography{aa_ACHer}

\end{document}